\documentclass[structabstract]{aa} 
\usepackage[round]{natbib}
\usepackage{graphicx}
\usepackage{txfonts}
\usepackage{aalongtable}
\usepackage{supertabular}

\bibpunct{(}{)}{;}{a}{}{,} % to follow the A&A style
\begin{document}
\title{Star-galaxy separation by far-infrared color-color diagrams for the AKARI FIS All-Sky Survey 
(Bright Source Catalogue Version $\beta$-1)
}
 
 \titlerunning{Star-galaxy separation by far infrared colors}
 \authorrunning{Pollo, Rybka, \& Takeuchi}

\author{A.\ Pollo \inst{1,2}, P.\ Rybka\inst{2}, T\ T.\ Takeuchi \inst{3}}%.

\institute{
	The Andrzej So{\l}tan Institute for Nuclear Studies, ul.\ Ho\.{z}a 69, 00-681 Warsaw, Poland
	\email{apollo@fuw.edu.pl}
	\and
	The Astronomical Observatory of the Jagiellonian University, ul.\ Orla 171, 30-244 Krak\'{o}w, Poland
	\and
Institute for Advanced Research, Nagoya University, Furo-cho, Chikusa-ku, Nagoya 464-8601, Japan
 }

\date{Received ; accepted }

% \abstract{}{}{}{}{} 
% 5 {} token are mandatory
 
  \abstract
	{}
{To separate stars and galaxies in the far infrared AKARI All-Sky Survey 
data, we have selected a sample with the complete color 
information available in the low extinction regions of the sky and constructed 
color-color plots for these data. We looked for the method to separate 
stars and galaxies using the color information.
}
{We performed an extensive search for the counterparts of these selected 
All-Sky Survey objects in the NED and SIMBAD databases. 
Among 5176 selected objects, we found {4272} galaxies, 
{382 other extragalactic objects,} {349 Milky Way stars, 50 other 
Galactic objects,} and {101} sources detected before in various wavelengths 
but of an unknown origin. {22} sources were left unidentified. 
Then, we checked colors of stars and galaxies in the far-infrared flux-color 
and color-color plots.
}
	{In the resulting diagrams, stars form two clearly separated clouds. 
One of them is easy to be distinguished from galaxies and allows for 
a simple method of excluding a large part of stars using the far-infrared data. 
The other {smaller} branch, overplotting galaxies, 
 {consists of stars known to have an infrared excess, like Vega 
and some fainter stars discovered by IRAS or 2MASS. 
The color properties of these objects in any case make them very difficult to 
distinguish from galaxies.} 
%in addition to a certain number of seemingly correctly identified stars, contains many stellar identifications which, after a careful examination, proved to be most probably an effect of source confusion. 
%This makes the method of star-galaxy separation by the FIR color-color plots even more powerful.
}
{We conclude that the FIR color-color diagrams allow for a high-quality 
star-galaxy separation. With the proposed simple method we can 
select more that 95~\% of galaxies rejecting {at least} 80~\% of stars.}
\keywords{
surveys -- dust, extinction -- galaxies: fundamental parameters -- infrared: galaxies -- infrared: stars
}

\maketitle
%
%________________________________________________________________

\section{Introduction}

Emission from galaxies at wavelengths beyond a few micrometers is produced mainly by dust.
As widely known, star formation activity is always accompanied by dust production,
{
though the actual mechanism of dust supply is not very well understood yet.
}
%Its source are for example, supernovae, which are the final phase of massive short-lived stars; they return 
%back metals, which they have produced at the stellar core, in the form of dust.
The energy emitted by massive stars at ultraviolet (UV) is scattered efficiently and
finally absorbed by dust grains.
The heated grains re-emit the energy at far-infrared (FIR).
Thus, radiative processes in FIR are related to the composition and the amount of dust in galaxies 
as well as to the properties of their stellar population, especially newly formed massive stars.
Many studies have shown that significant amount of star formation in galaxies is obscured by dust
\citep{lefloch05,caputi07,buat07a,buat07b,buat08,reddy08}.
Especially, such ``hidden'' star formation is revealed to become more and more important
with increasing redshift from $z = 0$ to 1 \citep{takeuchi05a}. 
Therefore, FIR observations have crucial importance to understand the {\it true} star formation
activity in the Universe.

Any FIR surveys would first provide us with a point source catalog, and extragalactic astrophysicists 
may want to have a reliable list of galaxy candidates from the FIR catalog. 
However, usually it is not a trivial task to classify detected sources and extract a specific class of 
objects only from FIR information.
Obviously not only galaxies would be detected at FIR, but also many other kinds of Galactic objects
would be included in 
{
a purely flux-limited catalog which is obtained as a first product of the survey.
}
Very crudely one may cut out the Galactic plane to avoid Galactic star-forming regions or H{\sc ii}
regions, but still there are numerous stars with dust, e.g., Vega-like stars and AGBs.
An effective and convenient method for star-galaxy separation which makes use of FIR information
only would be desired.
Such a method would also be very useful to construct a good candidate list of dusty stars.

The Infrared Astronomical Satellite \citep[IRAS; ][]{neugebauer84} has brought a vast amount of 
statistics and very efficient methods have been invented from IRAS Point Source Catalog (PSC). 
The four bands of IRAS enabled us to perform even a very detailed classification of extragalactic
and various galactic objects, like blue and red galaxies, Seyferts and QSOs, carbon stars, 
H{\sc ii} regions, reflection nebulae, planetary nebulae, T Tauri stars etc.
A {thorough} description of the method from IRAS can be found in \citep[e.g., ][]{walker89}.

After many years since IRAS, the advent of AKARI (ASTRO-F) opened a new window to 
explore the Universe, as a survey-oriented space telescope 
{at MIR and FIR} 
\citep{murakami07}.
The primary purpose of the mission is to provide second-generation infrared (IR) 
catalogs to obtain a better spatial resolution and a wider spectral coverage than 
the IRAS catalog.
AKARI is equipped with a cryogenically cooled telescope of 68.5~cm aperture diameter and two 
scientific instruments, the Far-Infrared Surveyor \citep[FIS; ][]{kawada07} and the
Infrared Camera \citep[IRC; ][]{onaka07}.
Among various astronomical observations performed by AKARI, an all sky survey with FIS 
{and IRC} has 
been carried out (AKARI All-Sky Survey). 

Since FIS is an instrument dedicated to FIR $\lambda = 50 \mbox{--} 180\;\mu$m, all the AKARI FIS 
bands are in the FIR wavelengths: {\it N60} ($65\;\mu$m), {\it WIDE-S} ($90\;\mu$m), 
{\it WIDE-L} ($140\;\mu$m), and {\it N160} ($160\; \mu$m) \citep{kawada07}. 
Hereafter, we use a notation $S_{65}$, $S_{90}$, $S_{140}$ and $S_{160}$ for 
flux densities in these bands.

Especially, since FIS has sensitivity at longer wavelengths than IRAS,  a new method of
classification scheme is needed if we try to select a list of a certain class of objects.
Such a scheme is not merely an empirical technique but also will provide us with 
a new understanding of objects with cool dust which were difficult to detect by IRAS
bands.

Since our central interest is on the physics of IR galaxies, we set our main aim to select galaxies 
by fluxes at four AKARI FIS bands.
In this paper, we present the first color-color and color-flux diagrams obtained by the All-Sky Survey data
only based on FIS bands,
and show the method of star-galaxy separation with these diagrams.

The paper is organized as follows: 
in Section~\ref{sec:data}, we present the data: the sample with the complete color information, 
In Sections~\ref{sec:flux_color} and \ref{sec:color_color}, we present the FIR flux-color and color-color 
diagrams of the data, respectively, and we show how galaxies and stars can be separated in them. 
We show our results and conclusions in Section~\ref{sec:conclusion}.

\section{The data}\label{sec:data}

\begin{figure}[t]
  \centering
  \includegraphics[angle=270,width=0.45\textwidth]{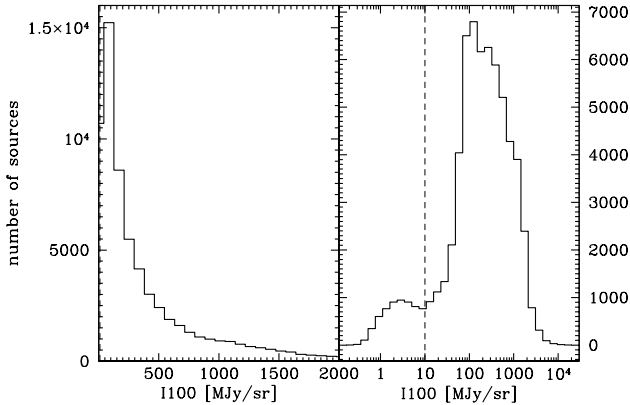}
  \caption{
Histogram of the number of objects as a function of the COBE DIRBE $100\;\mu$m 
sky emission intensity at the position of each object, made from \citep{schlegel98}.
The vertical dashed line in the right panel represents the criterion we have adopted
on the Galactic cirrus emission for this analysis ($I_{100} < 10\;\mbox{MJy}\,\mbox{sr}^{-1}$).
}\label{fig:hist}
\end{figure}

%\clearpage

\begin{table}[hb]
\caption{Classification of 5176 sources with a complete photometry located in the sky regions of low 100 $\mu$m emissivity {($I_{100} < 10\;\mbox{MJy sr}^{-1}$}) }
\label{tab:tabgal}
\centering 
\begin{tabular} {llc}
\hline
\hline
Galaxies & & {4272} \\ %4232 \\
\hline
{Other e}xtragalactic objects & & {382} \\ %213 \\
\hline
& pairs of galaxies & 65 \\
& triplets of galaxies & 4 \\
& groups of galaxies & 6 \\
& parts of galaxies & {50} \\
& stars in nearby galaxies & {148} \\ %134 \\
& {\sc Hii} regions & 7{6} \\ %6 \\
& supernovae hosts & 28 \\
& quasi-stellar objects & 2 \\
& emission-line objects & 2 \\
%& other extragalactic sources & 47 \\
%& radio sources & 7 \\
%& infrared sources & 25 \\
%& visual sources & 3 \\
%& UV sources & 1 \\
%& X-ray sources & 11 \\
\hline
\hline
%Stellar objects & & 545 \\
{Milky Way stars} & & {349} \\
\hline
& variable stars & {264} \\ % 246 \\
%& stars & 196 \\
& peculiar stars & {38} \\ %52 \\
& {other stars} & {47} \\
\hline
{Other Galactic objects} & & {50} \\
& associations of stars & {2} \\
& planetary nebulae & 35 \\
& reflection nebulae & 2 \\
& other nebulae & 1 \\
& {Galactic cirruses} & {5} \\
&{{\sc Hii} regions} & {1} \\
& OH masers & 4 \\
\hline
\hline
Other sources & & 101 \\
\hline
radio sources & & {33} \\ %4 \\
infrared sources & & 48 \\
X-ray sources & & 20 \\
\hline
\hline
Not identified sources & & {22} \\ % 30 \\
\hline
\hline
\end{tabular}
\end{table}

%\clearpage
As mentioned in Introduction, one of the main missions of AKARI was to carry 
out the All-Sky Survey in the four photometric bands in the far-infrared 
wavelength ranging in 50--180~$\mu$m with FIS \citep{kawada07}. 
The FIS scanned 94 percent of the entire sky more than twice in the 16 months of 
the cryogenic mission phase.

The AKARI FIS Bright Source Catalogue (hereafter BSC) is the first primary 
catalog from the AKARI All-Sky Survey. 
Data from the version $\beta$-1 of this catalog are used in this work. 
AKARI BSC is supposed to have a uniform detection limit, corresponding to per scan 
sensitivity, over the entire sky, except for very bright sky parts where different 
data acquisition mode had to be applied. 
Redundant observations were used to increase the reliability of the detection.  vs.\ 
A summary of the All-Sky Survey can be found in \citet{yamamura09}.

The AKARI FIS BSC provides data for 64311 sources, among them 43342 sources with fluxes 
measured at all the four FIR bands. 
For each detected source, AKARI source identifier, equatorial coordinates of the 
source position and flux densities in the four FIR bands are given. 
The position accuracy of the FIS BSC is $8''$, since the source extraction is made 
with grids of this size.
Effective size of the point spread function of AKARI FIS in FWHM is estimated to be $37 \pm 1''$,
$39 \pm 1$, $58 \pm 3''$, and  $61 \pm 4''$ at {\it N60}, {\it WIDE-S}, {\it WIDE-L}, and
{\it N160}, respectively \citep{kawada07}.
Errors are not estimated for each individual source, but instead they are in total 
estimated to be 35~\%, 30~\%, 60~\%, and 60~\% at {\it N60}, {\it WIDE-S}, 
{\it WIDE-L}, and {\it N160}, respectively \citep{yamamura08}.

\begin{figure}[t]
  \centering
  \includegraphics[angle=90,width=0.45\textwidth]{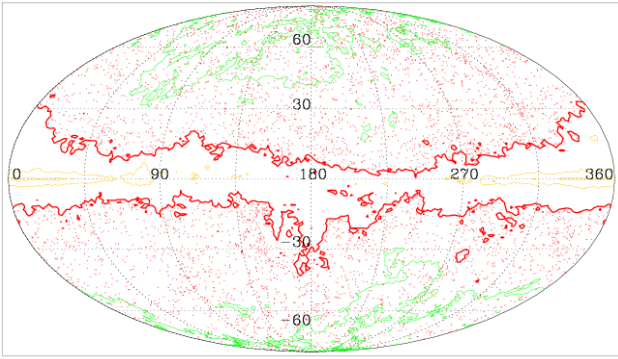}
  \caption{
	{The sky distribution of the AKARI FIS sources in the low FIR sky emissivity 
	regions, overplotted with the COBE DIRBE $100\;\mu$m sky emission intensity 
	$I_{100}$. 
	Contour levels are logarithmically spaced from $I_{100} = 10^{-1}, 10^{0}, 10^{1}, 10^{2}, 
	10^{3}, \mbox{ and } 10^{4}\;\mbox{MJy sr}^{-1}$. 
	The thick contours indicate our criterion $I_{100} < 10\;\mbox{MJy sr}^{-1}$.
	}
	}\label{fig:sky_dist}
\end{figure}

%\clearpage
The Galactic $100\;\mu$m emission at each position was obtained from the so-called 
SFD maps \citep{schlegel98}. 
The distribution of the number of the sources as a function of the 
100-$\mu$m Galactic emission at the position of each source is shown in Fig.~\ref{fig:hist}: 
left panel is the distribution in linear scale, 
and right panel is in logarithmic scale.
For optical observations, as widely known, heavily obscured regions of the sky 
are not ideal for source finding and photometry.
Since the biggest available database is often found at {\it B}-band, the extinction
would be large and may affect the probability of counterpart association, even if
we can correct the extinction for detected objects.
At FIR side, dust grains in the Galactic interstellar medium emit thermal radiation 
at these wavelengths.
Quality of flux measurement at high-Galactic foreground regions would be 
affected by the FIR diffuse Galactic emission, and sometimes it is difficult to remove
the effect completely.
Since both stem from the Galactic dust, putting a threshold on the extinction 
$E(B-V)$ and the diffuse foreground radiation would give almost 
equivalent result.
Hence, we have selected objects lying in regions 
associated with low Galactic emission at 100~$\mu$m.
We set our threshold to be $10\;\mbox{MJy\,sr}^{-1}$, which is a very conservative 
value, but thanks to this, we can perform the subsequent analysis safely. 
This step left us with 6030 sources.
The sky distribution of these sources is presented in Fig.~\ref{fig:sky_dist}.
By this criterion, the selected sources naturally avoid the Galactic plane region.
Hence, most of the Galactic sources which are the majority of the AKARI FIS sources
are not included in our sample. 
This is seen in the left panel of Fig.~\ref{fig:hist}.
The number of the sources in the selected regions is much smaller than that of
the excluded regions.
However, the area of the selected region is $77.5\;\%$ of the total sky area.

Among selected sources, 5176 sources have full four-band information.
These sources were then used to construct color-color and flux-color 
diagrams presented in the subsequent sections.

The next step was to identify thus selected sources with known objects in public databases. 
For that purpose, we searched the NED and SIMBAD databases looking for nearest 
counterparts within $40''$-circle around positions of AKARI sources.
This diameter of $40''$ was chosen since the FWHM of AKARI FIS at
{\it WIDE-S} is $39''$, i.e., this is large enough compared to the position
accuracy of AKARI ($8''$) and small enough not to have too many interlopers.
The summary of the identified sources is given in Table \ref{tab:tabgal}. 
In total, {399} Galactic and {4654} extragalactic objects were identified. 
In addition, we found {101} sources observed in other wavelengths for 
which it is not 
clear whether they are of Galactic or extragalactic origin. 
Only {22} sources were left unidentified.

Though we do not have direct information of the redshift distribution of 
the sample, there are related studies which can provide us with an idea:
\citet{takeuchi10} associated the FIS BSC with SDSS and IRAS PSC$z$
and showed the redshift distribution of the subsample of BSC with
ultraviolet, optical, NIR, and FIR counterparts. 
They found that the vast majority of their sample locate at $z < 0.05$
(see their Fig.~3).
\citet{malek10} made a counterpart association of the 90-$\mu$m
sources detected in the deep FIS survey data in the AKARI Deep 
Field-South.
The fraction of redshift association was not sufficiently high, but 
\citet{malek10} found that most of bright galaxies ($S_{90} > 100\;
\mbox{mJy}$) are at $z \leq 0.01$.
These studies suggest that most of our sample locate at very low redshifts.
Hence, we can safely ignore any cosmological effects such as $K$-correction,
since it is much smaller than the uncertainties in flux measurements.

{ 
%We comment on a possible effect of redshift on this analysis for further
%studies.
For further studies, however, a possible effect of redshift may become
important.
Since our sample is selected by FIS {\it WIDE-S} ($90\;\mu$m), it
is dominated by galaxies with relatively cool dust.
Hence, if we had a high-redshift galaxy in this sample, it would appear 
to be redder on the flux-color or color-color diagrams. Then,
in the diagrams we discuss in this paper, such a galaxy would be even 
further from stars than nearby galaxies. However, the effect is not 
very drastic because the FIS does not have submillimeter bands.
This effect was discussed in order to select high-redshift galaxies from
the FIS four bands by \citet{takeuchi99} (see their Fig.~9).
}

\section{Flux-color diagrams}\label{sec:flux_color}

\begin{figure}[t]
	\centering
	\includegraphics[width=0.45\textwidth]{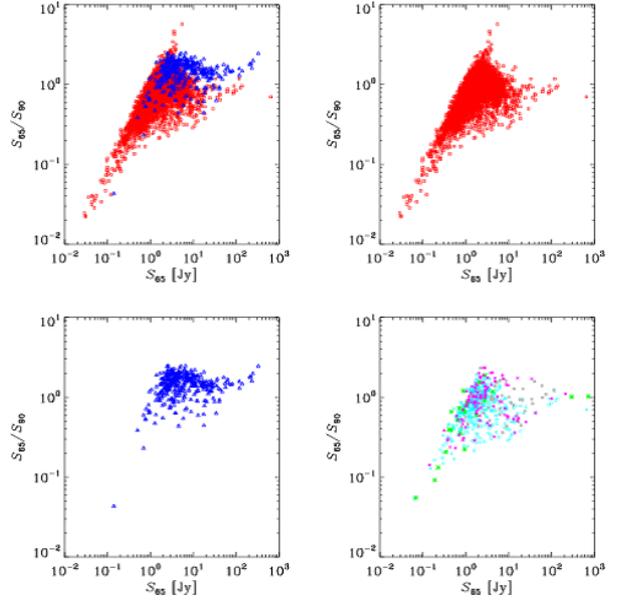}
	\caption{The $S_{65}$--$S_{65}/S_{90}$ flux-color diagram.
	{\it Top-left}: galaxies and stars, 
	{\it top-right}: galaxies, 
	{\it bottom-left}: stars,
	and
	{\it bottom-right}: remaining objects: Galactic, 
   extragalactic, the ones which are with the identifications but we could 
	not decide whether they are Galactic or extragalactic, or ones which were 
	not identified.
	Symbols represent the sources as follows:
	open squares {(red in the onlnie version)}: galaxies;
    open triangles {(blue)}: the Galactic stars;  
	open circles {(cyan)}: other extragalactic objects (QSOs, stars in extragalactic objects);
	filled squares {(gray)}: other Galactic objects (star associations, nebulae, H{\sc ii} regions, cirruses);
	asterisks {(green)}: unknown origin;	
	crosses {(magenta)}: sources without counterparts.
    }\label{fig:fc1}
\end{figure}

\begin{figure}[t]
	\centering
	\includegraphics[width=0.45\textwidth]{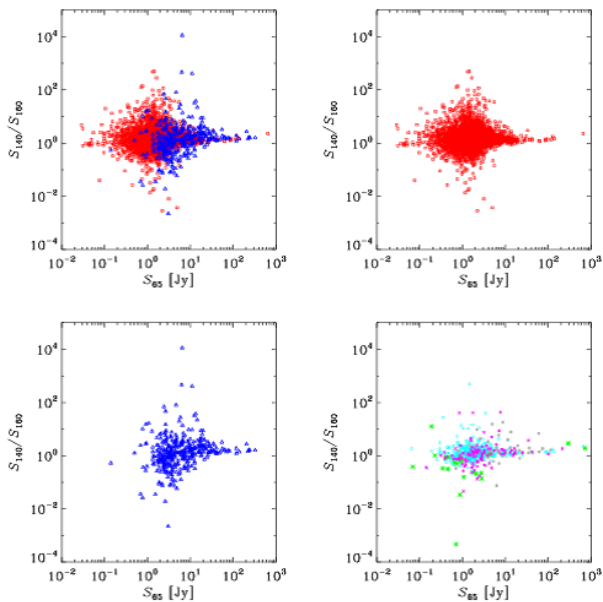}
	\caption{Same as Fig.~\ref{fig:fc1} but for $S_{65}$--$S_{140}/S_{160}$.
	}\label{fig:fc2}
\end{figure}
%Six flux-color diagrams constructed from four FIS measurements are
%presented in Figs.~\ref{fig:fc1}--\ref{fig:fc6}. 
{
Two examples of flux-color diagrams constructed from four FIS measurements are
presented in Figs.~\ref{fig:fc1} and \ref{fig:fc2}. 
Since other diagrams contain very similar information, remaining diagrams 
are presented only in electronic form, as online 
Figs.~\ref{fig:afc1}--\ref{fig:afc11}.
}

At {the} first glance, in the presented diagram{s} 
we clearly see two separate 
branches at brighter flux densities in Fig.~\ref{fig:fc1}:
one branch is for galaxies, and the other is for stars.

For Fig.~\ref{fig:fc2}, since both bands are longer than $100\;\mu$m,
it is very possible that both are at the Rayleigh--Jeans regime of their
dust emission.
Consequently, we cannot expect a clear distinction between stars
and galaxies because they are quite similar to each other.

For the other colors, the contiguous set of AKARI FIS bands \citep{kawada07}
enables us to grasp the peak of dust spectrum.
This reflects the distinct two branches on the flux-color diagrams.
Though stars show a slight double-peaked distribution, one branch 
is much more prominent than the second one. We revisit the issue of
the identities of these stars and, in particular, the smaller redder branch
of stars, later.

In sum, these diagrams encourage us to go further to explore the classification 
scheme from the AKARI FIS bands.

\section{Color-color diagrams}\label{sec:color_color}

\begin{table}[hb]
\caption{Division lines between stars and galaxies for color-color diagrams}
\label{tab:cclines}
\centering
\begin{tabular} {llrr}
\hline
\hline
Color 1 & Color 2 & $a$ & $b$ \\
\hline 
$S_{65}/S_{90}$ & $S_{65}/S_{140}$ & $-0.17$ & $0.33$ \\
$S_{65}/S_{90}$ & $S_{65}/S_{160}$ & $-4.09$ & $1.19$ \\
$S_{65}/S_{90}$ & $S_{90}/S_{140}$ & $-1.24$ & $0.32$ \\
$S_{65}/S_{90}$ & $S_{90}/S_{160}$ & $-4.09$ & $0.97$ \\
$S_{65}/S_{90}$ & $S_{140}/S_{160}$ & $-49.09$ & $10.01$ \\
$S_{65}/S_{140}$ & $S_{65}/S_{160}$ & $-2.95$ & $1.39$ \\
$S_{65}/S_{140}$ & $S_{90}/S_{140}$ & $-2.19$ & $1.20$ \\
$S_{65}/S_{140}$ & $S_{90}/S_{160}$ & $-2.46$ & $1.00$ \\
$S_{65}/S_{140}$ & $S_{140}/S_{160}$ & $-1.81$ & $0.60$ \\
$S_{65}/S_{160}$ & $S_{90}/S_{140}$ & $-0.46$ & $0.40$ \\
$S_{65}/S_{160}$ & $S_{90}/S_{160}$ & $2.36$ & $-1.04$ \\
$S_{65}/S_{160}$ & $S_{140}/S_{160}$ & $ 1.31$ & $-0.49$ \\
$S_{90}/S_{140}$ & $S_{90}/S_{160}$ & $-3.94$ & $1.38$ \\
$S_{90}/S_{140}$ & $S_{140}/S_{160}$ & $-9.20$ & $2.34$ \\
$S_{90}/S_{160}$ & $S_{140}/S_{160}$ & $1.19$ & $-0.30$ \\
\hline
\end{tabular}
\end{table}

\begin{table*}
\caption{The quality of star-galaxy separation in color-color diagrams}
\label{tab:separationgs}
\centering
\begin{tabular} {lcccccccc}
\hline
\hline
Diagram & \multicolumn{2}{c}{Galaxies in the galaxy cloud} & \multicolumn{2}{c}{Stars in the star cloud} & \multicolumn{2}{c}{Galaxies in the star cloud} & \multicolumn{2}{c}{Stars in the galaxy cloud} \\
 & number & percentage & number & percentage & number & percentage & number & percentage \\
\hline
 $S_{65}/S_{90}$ vs. $S_{65}/S_{140}$ & 4114 & 96.30 & 290 & 83.09 & 158 & 3.70 & 59 & 16.91 \\ 
 $S_{65}/S_{90}$ vs. $S_{65}/S_{160}$ & 4110 & 96.21 & 226 & 64.76 & 162 & 3.79 & 123 & 35.24 \\ 
 $S_{65}/S_{90}$ vs. $S_{90}/S_{140}$ & 4096 & 95.88 & 295 & 84.53 & 176 & 4.12 & 54 & 15.47 \\ 
 $S_{65}/S_{90}$ vs. $S_{90}/S_{160}$ & 4085 & 95.62 & 245 & 70.20 & 187 & 4.38 & 104 & 29.80 \\ 
 $S_{65}/S_{90}$ vs. $S_{140}/S_{160}$ & 4109 & 96.18 & 144 & 41.26 & 163 & 3.82 & 205 & 58.74 \\ 
 $S_{65}/S_{140}$ vs. $S_{65}/S_{160}$ & 4120 & 96.44 & 294 & 84.24 & 152 & 3.56 & 55 & 15.76 \\ 
 $S_{65}/S_{140}$ vs. $S_{90}/S_{140}$ & 4162 & 97.43 & 274 & 78.51 & 110 & 2.57 & 75 & 21.49 \\ 
 $S_{65}/S_{140}$ vs. $S_{90}/S_{160}$ & 4081 & 95.53 & 304 & 87.11 & 191 & 4.47 & 45 & 12.89 \\ 
 $S_{65}/S_{140}$ vs. $S_{140}/S_{160}$ & 4018 & 94.05 & 301 & 86.25 & 254 & 5.95 & 48 & 13.75 \\ 
 $S_{65}/S_{160}$ vs. $S_{90}/S_{140}$ & 4107 & 96.14 & 293 & 83.95 & 165 & 3.86 & 56 & 16.05 \\ 
 $S_{65}/S_{160}$ vs. $S_{90}/S_{160}$ & 4128 & 96.63 & 198 & 56.73 & 144 & 3.37 & 151 & 43.27 \\ 
 $S_{65}/S_{160}$ vs. $S_{140}/S_{160}$ & 4147 & 97.07 & 288 & 82.52 & 125 & 2.93 & 61 & 17.48 \\ 
 $S_{90}/S_{140}$ vs. $S_{90}/S_{160}$ & 4094 & 95.83 & 286 & 81.95 & 178 & 4.17 & 63 & 18.05 \\ 
 $S_{90}/S_{140}$ vs. $S_{140}/S_{160}$ & 4098 & 95.93 & 289 & 82.81 & 174 & 4.07 & 60 & 17.19 \\ 
 $S_{90}/S_{160}$ vs. $S_{140}/S_{160}$ & 4081 & 95.53 & 291 & 83.38 & 191 & 4.47 & 58 & 16.62 \\ 
\hline
\hline
\end{tabular}
\end{table*}

\begin{table*}
\caption{{The quality of star-galaxy separation in color-color diagrams {\it outside} of the areas between the lines indicating photometric errors}}
\label{tab:separationgs_minus}
\centering
\begin{tabular} {lcccccccccc}
\hline
\hline
Diagram & \multicolumn{2}{c}{Total number} & \multicolumn{2}{c}{Galaxies} & \multicolumn{2}{c}{Stars} & \multicolumn{2}{c}{Galaxies} & \multicolumn{2}{c}{Stars} \\  
& \multicolumn{2}{c}{} & \multicolumn{2}{c}{in the galaxy cloud} & \multicolumn{2}{c}{in the star cloud} & \multicolumn{2}{c}{in the star cloud} & \multicolumn{2}{c}{in the galaxy cloud} \\ 
& galaxies & stars & number & percentage & number & percentage & number & percentage & number & percentage \\
\hline
 $S_{65}/S_{90}$ vs. $S_{65}/S_{140}$ & 3328 & 157 & 3297 & 99.07 & 142 & 90.45 & 31 & 0.93 & 15 & 9.55 \\ 
 $S_{65}/S_{90}$ vs. $S_{65}/S_{160}$ & 2989 & 67 & 2970 & 99.36 & 38 & 56.72 & 19 & 0.64 & 29 & 43.28 \\ 
 $S_{65}/S_{90}$ vs. $S_{90}/S_{140}$ & 3370 & 185 & 3335 & 98.96 & 168 & 90.81 & 35 & 1.04 & 17 & 9.19 \\ 
 $S_{65}/S_{90}$ vs. $S_{90}/S_{160}$ & 2583 & 52 & 2568 & 99.42 & 34 & 65.38 & 15 & 0.58 & 18 & 34.62 \\ 
 $S_{65}/S_{90}$ vs. $S_{140}/S_{160}$ & 2123 & 31 & 2118 & 99.76 & 0 & 0.00 & 5 & 0.24 & 31 & 100.00 \\ 
 $S_{65}/S_{140}$ vs. $S_{65}/S_{160}$ & 3631 & 207 & 3600 & 99.15 & 189 & 91.30 & 31 & 0.85 & 18 & 8.70 \\ 
 $S_{65}/S_{140}$ vs. $S_{90}/S_{140}$ & 3901 & 165 & 3865 & 99.08 & 138 & 83.64 & 36 & 0.92 & 27 & 16.36 \\ 
 $S_{65}/S_{140}$ vs. $S_{90}/S_{160}$ & 3491 & 226 & 3444 & 98.65 & 213 & 94.25 & 47 & 1.35 & 13 & 5.75 \\ 
 $S_{65}/S_{140}$ vs. $S_{140}/S_{160}$ & 2896 & 181 & 2855 & 98.58 & 170 & 93.92 & 41 & 1.42 & 11 & 6.08 \\ 
 $S_{65}/S_{160}$ vs. $S_{90}/S_{140}$ & 3525 & 201 & 3489 & 98.98 & 184 & 91.54 & 36 & 1.02 & 17 & 8.46 \\ 
 $S_{65}/S_{160}$ vs. $S_{90}/S_{160}$ & 3129 & 59 & 3111 & 99.42 & 37 & 62.71 & 18 & 0.58 & 22 & 37.29 \\ 
 $S_{65}/S_{160}$ vs. $S_{140}/S_{160}$ & 3353 & 142 & 3335 & 99.46 & 129 & 90.85 & 18 & 0.54 & 13 & 9.15 \\ 
 $S_{90}/S_{140}$ vs. $S_{90}/S_{160}$ & 3345 & 157 & 3308 & 98.89 & 142 & 90.45 & 37 & 1.11 & 15 & 9.55 \\ 
 $S_{90}/S_{140}$ vs. $S_{140}/S_{160}$ & 2787 & 99 & 2758 & 98.96 & 93 & 93.94 & 29 & 1.04 & 6 & 6.06 \\ 
 $S_{90}/S_{160}$ vs. $S_{140}/S_{160}$ & 2245 & 80 & 2223 & 99.02 & 76 & 95.00 & 22 & 0.98 & 4 & 5.00 \\ 
\hline
\hline
\end{tabular}
\end{table*}

\begin{figure*}[t]
  \centering
  \includegraphics[width=0.8\textwidth,clip]{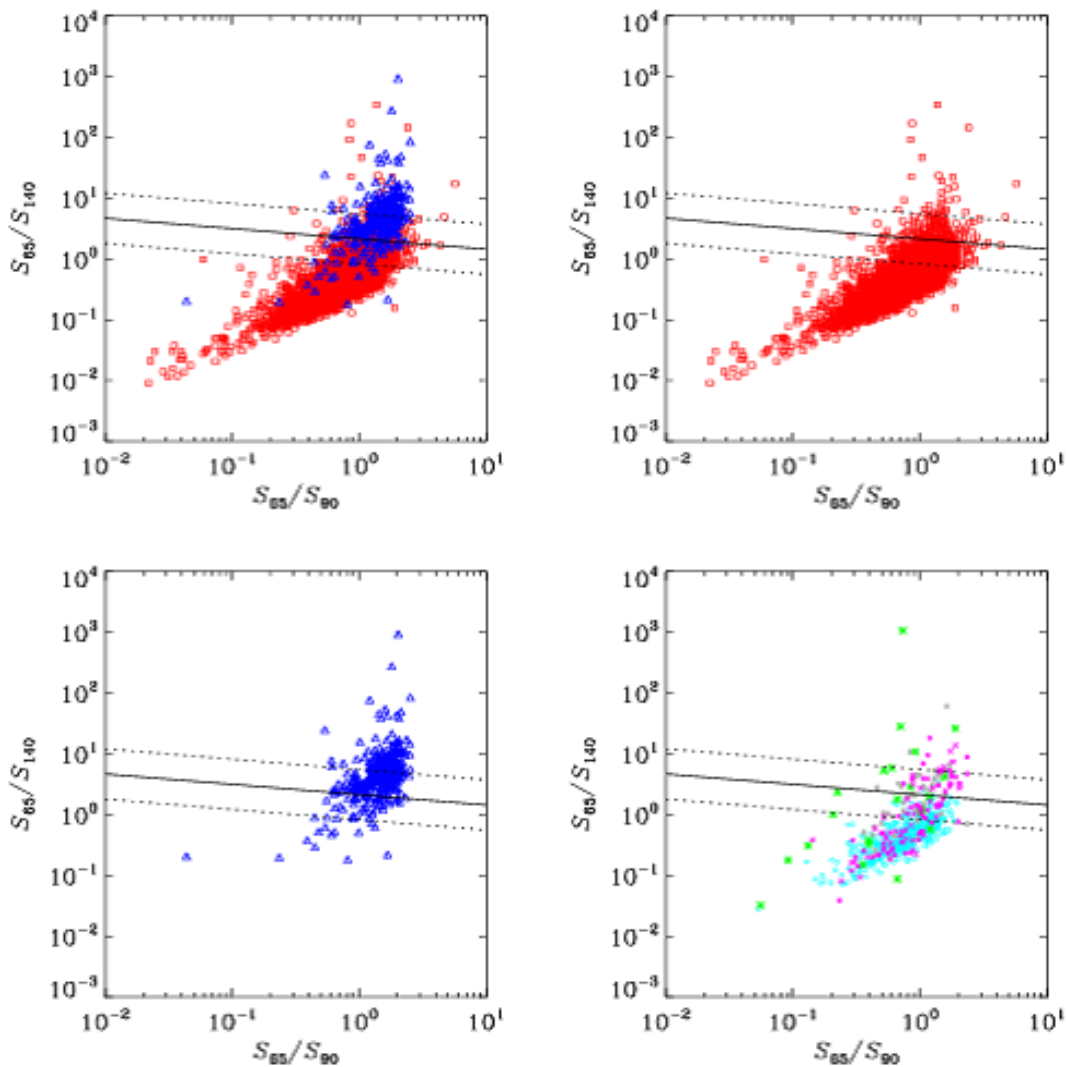}
  \caption{The $S_{65}/S_{90}$--$S_{65}/S_{140}$ color-color diagrams of the sample.
	The format and symbols in the panels are the same as those of flux-color diagrams.
}\label{fig:cc1}
\end{figure*}

\begin{figure*}[t]
	\centering
	\includegraphics[width=0.8\textwidth]{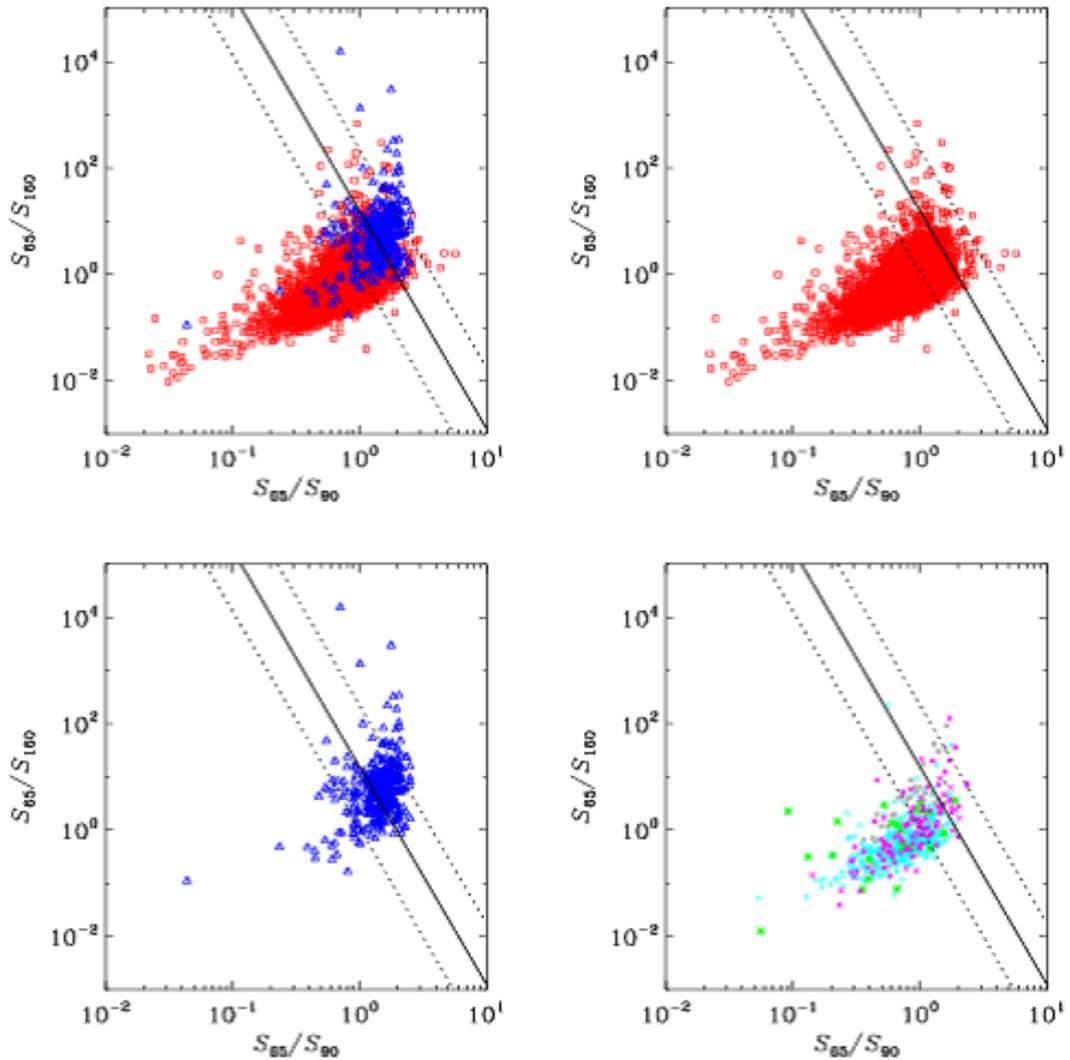}
	\caption{Same as Fig.~\ref{fig:cc1} but for $S_{65}/S_{90}$--$S_{65}/S_{160}$.}  
	 \label{fig:cc2}
\end{figure*}
 
\begin{figure*}[t]
	\centering
	\includegraphics[width=0.8\textwidth]{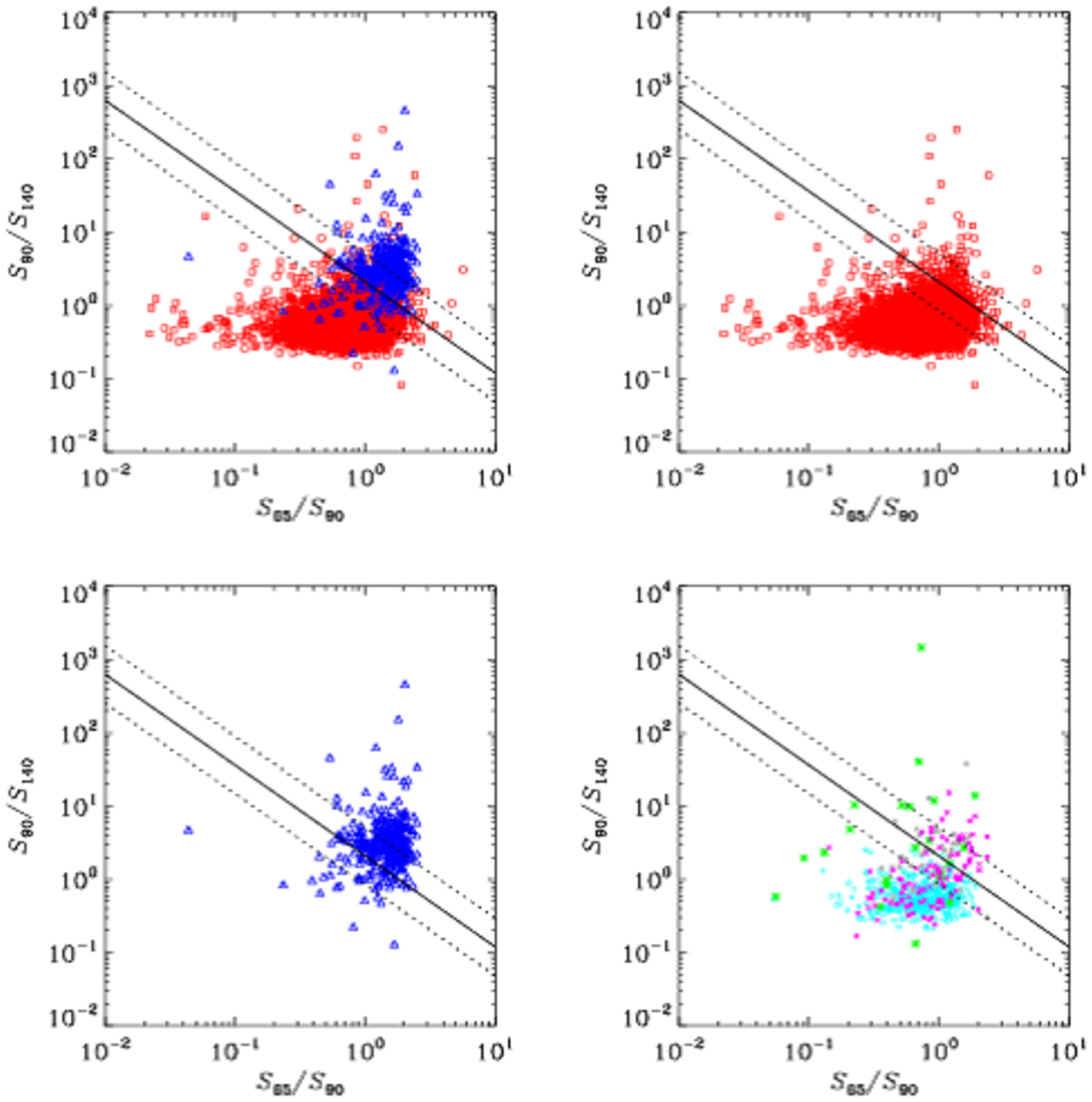}
	\caption{Same as Fig.~\ref{fig:cc1} but for $S_{65}/S_{90}$--$S_{90}/S_{140}$.}  
	 \label{fig:cc3}
\end{figure*}

\begin{figure*}[t]
	\centering
	\includegraphics[width=0.8\textwidth]{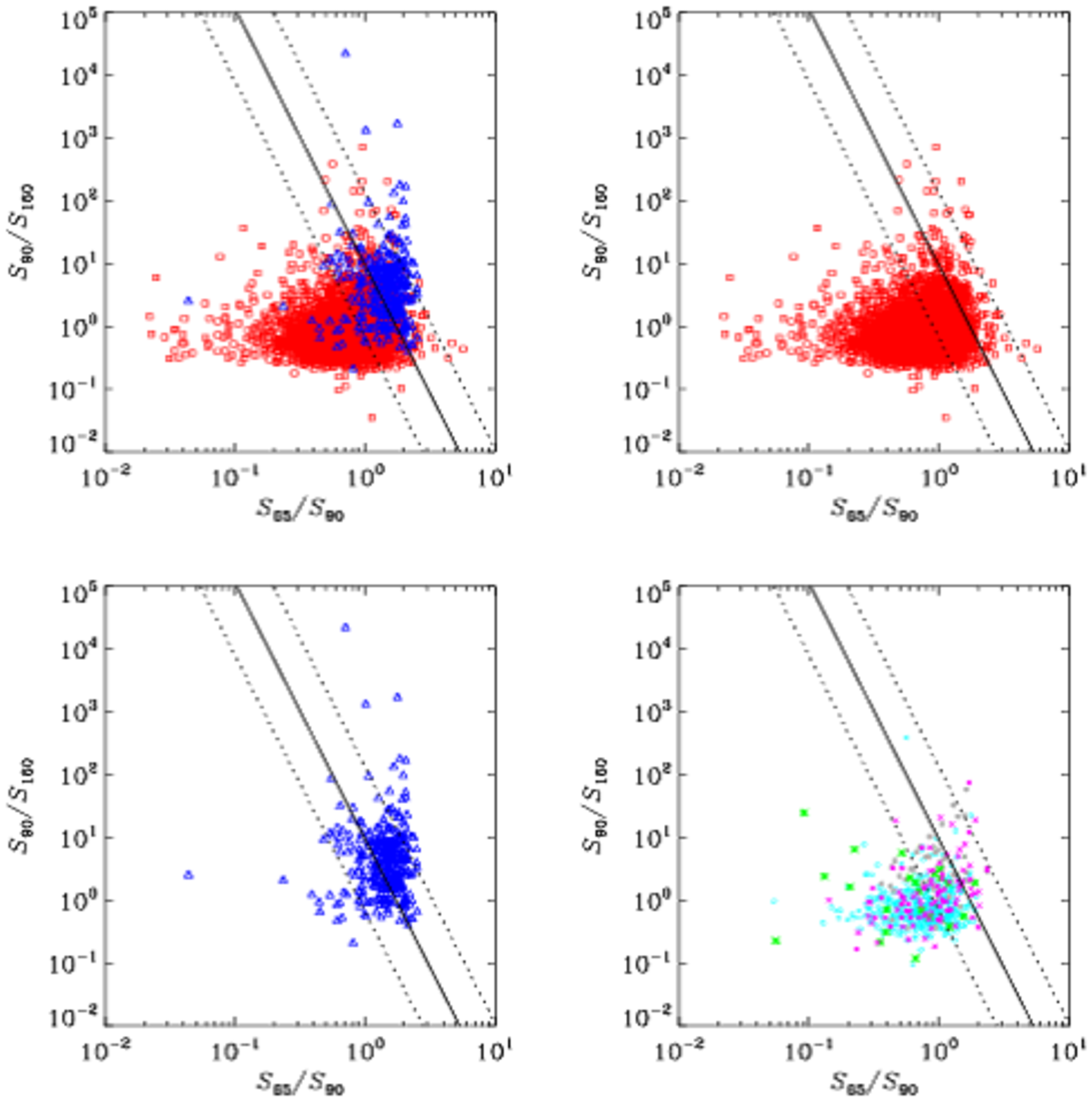}
	\caption{Same as Fig.~\ref{fig:cc1} but for $S_{65}/S_{90}$--$S_{90}/S_{160}$.}  
	 \label{fig:cc4}
\end{figure*}

We show the color-color diagrams constructed from four FIS measurements
in Figs.~\ref{fig:cc1}--\ref{fig:cc4}.
Again, we do not show all, but four among 15 independent diagrams.
The rest of them is presented only in an electronic form as online Figs.~\ref{fig:cc5}--\ref{fig:cc15}. 

Based on these scatter plots, we decided the division line between stars and galaxies.
In this work, we did not try to make a detailed statistical discriminant analysis 
but put them by hand.
One may have some anxiety that this procedure would be ``subjective'', but in the 
following examination we show 
that our determination works very well and is almost perfectly consistent.
We plan to show a more statistically sophisticated analysis elsewhere.

The central solid straight line shown in each panel represents the best 
division boundary 
separating galaxies and stars. 
The parameters of these lines for each diagram, $a$ and $b$ in the following
equation
\begin{equation}
  \mbox{Color 2} = a \times \left( \mbox{Color 1} \right) + b \;. 
\end{equation} 
are listed in Table~\ref{tab:cclines}. 
The dotted lines represent
the uncertainty ranges in the color determination corresponding to the average
photometric errors: 35~\%, 30~\%, 60~\%, and 60~\% at {\it N60}, {\it WIDE-S},
{\it WIDE-L}, and {\it N160}, respectively.
As shown in Table~\ref{tab:separationgs}, 
in all cases the quality of separation is very good, especially for galaxies. 
{}To two sides of 
the dividing line we refer there as to the galaxy cloud (the side dominated
by galaxies) and the stellar cloud (the side dominated by stars). It is 
shown that galaxies are selected 
with the accuracy ranging between 95~\% and 98~\%. Stars, again, form two 
separate branches among which one, {usually consisting of more than
80~\% of stars}, is well separated and the second, smaller one 
overlaps galaxies. 

The quality of the division is affectedby the photometric errors, 
at least in case of some diagrams.
As shown in Table~\ref{tab:separationgs_minus},
taking into account {\it only} objects located in the diagrams {\it outside} 
the ``error bands'' between the lines corresponding to photometric errors, 
we obtain a much better separation of galaxies, always above 98~\%, and often
well above 99~\%. 
In contrast, the quality of the separation of stars becomes less reliable.
First, the number statistics of stars often becomes poor
with such a selection. 
Second, we can observe a clear split between
diagrams: in most cases the quality of star separation becomes better,
above 90~\%, which can advocate that both our division and the error 
estimation are correct. 
However, in some cases (the ones with the smallest
numbers of stars left outside the error bands) the quality estimated
in such a way drops down significantly. Apart of the statistical reason,
this fact may be related to larger uncertainties in 
the measurement of the FIR fluxes, in particular $S_{160}$.

In this place it is also worth drawing reader's attention to 
Figs.~\ref{fig:cc1} and \ref{fig:cc2}. 
The difference between them is the usage
of color along the $y$ axis: in Fig.~\ref{fig:cc1} it is $S_{65}/S_{140}$
and in Figure \ref{fig:cc2} - $S_{65}/S_{160}$. 
Given the small difference in wavelengths, we expect that for a given 
object these colors and the separation lines would be similar on the diagrams. 
However, the best division lines are significantly different from each other. 

With the division lines applied, the quality of the galaxy separation is 
in both these diagrams very good, but the quality of star separation is 
significantly lower in case of Fig.~\ref{fig:cc2}. 
A plausible explanation, again, may be uncertainties larger than
the assumed ones in the flux measurement at {\it N160} 
in case of the $\beta$-1 version of the FIS PSC.
It should be stressed, however, that even taking this possibility 
into account, the quality of the separation in all the diagrams remains high.

As shown in Table~\ref{tab:consistency}, the consistency of the division
is very high for selected galaxies (usually higher than 90~\%). 
In case of stars,
the consistency of the division
varies much more strongly among flux-color diagrams, but in most cases 
the percentage of correct categorization is also very high. 

We checked the properties of stars belonging to both groups.
% (I.\ Yamamura, private communication). 
We used the separation obtained in 
the diagram $S_{65}/S_{90}$ vs.\ $S_{90}/S_{140}$,
which is shown in Fig.~\ref{fig:cc3}. 
{}From the consistency checks
we can assume that the diagnosis of these stars will be roughly applicable 
for most of the other diagrams as well. 

A well separated branch of stars in this diagram consists of 29{5} objects.
Most of them are optically bright stars. 
This sample is dominated by variable stars, mainly Mira-type stars 
(35~\%), and other pulsating stars (40~\%). There are also a
few binary systems.
Among the rest we find evolved stars like carbon stars (6~\%), stars 
with envelope of OH/IR type (4~\%), a few other AGB and post-AGB 
stars. We find also a few faint sources identified as stars by big
surveys like SDSS.
We call this branch of stars ``Branch I''.

The second {less prominent} branch of stars, mixed with galaxies,
consists of {54} objects.
We call this branch ``Branch II''.

%A more careful examination of these objects 
%revealed that only 47 among these objects are known bright stars.
{A significant fraction of the objects in this branch are stars
which were originally discovered in the infrared by 
IRAS or 2MASS (28~\%) or other stars already known to have 
a significant excess of the infrared luminosity (most notable
of them is Vega). 
In addition, we have again a significant number
of different types of pulsating variables (32~\%) but this time only
two Mira-type stars among them, five faint stars and three T Tauri stars.}

It is also interesting to take a look at the distribution of the 
other Galactic sources: they are divided much more equally
between two branches, and slightly more (60~\%) overlaps for Branch II. 
All the masers, as well as 43~\% of planetary nebulae are located the area
corresponding to Branch I. The remaining 57~\% of planetary nebulae,
as well as all the other nebulae, an {\sc Hii} region and all the Galactic
cirruses overlap Branch II.

For that reason, if make a division not just between stars 
and galaxies, but between the galactic and extragalactic objects, the
percentage of the Galactic objects in the galaxy cloud becomes
typically a few percent higher.

As we could see, stars and other Galactic objects that occupy the area 
of the ``galaxy cloud'' typically belong to special classes
of extremely red sources and it may require much more sophisticated
techniques than just simple one-line division in the color-color
diagram to distinguish them from the extragalactic sources.

In sum, the division we presented and discussed here, allows for
an efficient separation of more than 95~\% of galaxies and around 80~\%
of stars in the two parts of the diagram. 
As it was already 
mentioned before, some diagrams, even if still applicable, are obviously 
worse than the others, in particular when it comes to the separation
of stars. A plausible reason are larger uncertainties in the
FIR flux measurement of the beta-1 version of the FIS PSC, 
most probably in the longest wavelength filter {\it N160}.

A more detailed classification of Galactic and extragalactic sources
based
on their color properties will be a subject of a subsequent paper.

Sources for which it is unknown whether their origin is Galactic or
extragalactic in the diagrams appear on both sides of the dividing lines.
While only about 10~\% of a total number of objects from our initial sample
is located at the ``stellar'' part of the diagrams, for these
objects the fraction in the stellar cloud raises to about {3}0~\%.
One possible interpretation is that these objects are more often of
a Galactic origin.
However, they may also be
extragalactic sources with extreme colors.

Objects for which no counterpart was found (there are only {22} 
of them) are also distributed in both parts of diagrams {also more
than 30~\% of them are located in the ``stellar'' part of the diagrams}.
%Around one third of 
%them is located distinctly far from all the identified objects. 
%They may be an effect of mistaken photometry. 
%However, it cannot be excluded that they are real objects with extreme 
%colors. 
Their identity will be investigated in future works.

\section{Conclusions}\label{sec:conclusion}

\begin{enumerate}
\item The far infrared flux-color diagrams show two clear branches 
of sources. 
One of them is populated almost only by stars, the second one is dominated 
by galaxies. 
\item The far infrared color-color diagrams allow for a high quality 
star-galaxy separation. In all the combinations of color-color diagrams 
we can distinguish two separate clouds. 
It was shown that one of them contains in all 
cases more than 9{5}~\% of galaxies and the other one 
{in most cases} consists in more than {8}0~\% of stars. 
\item At a first glance the population of stars is divided into 
(roughly) two categories: a {larger group} occupies a separate area on the diagrams while the 
{remaining smaller part} resides in the ``galaxy cloud''. 
A more careful examination has revealed
that stars occupying a separate area on the color-color diagram are
{mostly} known bright stars. { This group is dominated by
pulsating variable stars, among them many Mira-type variables, and
contains also a number of evolved stars.} 
{Branch II of stars, overlapping the galaxy cloud, includes
stars discovered by the infrared surveys, bright stars known to have
an infrared excess, like Vega, pulsating stars, a few T Tauri stars and
faint red stars.}
%In contrast, around 70~\% of stellar identifications in the smaller redder 
%branch are most probably an effect of contamination. 
%Taking this observation into account, 
We estimate that the percentage 
of galaxies intervening the stellar part of the diagram is in each case 
less than 6~\%, and the percentage of stars in the galaxy part of 
the diagram - {in most cases} less than 20~\%. {A few 
exceptions can be probably attributed to some systematic error in
the flux estimation, in particular in the $N160$ band.}
{
Obviously, since this analysis was made in the sky regions with
low-Galaxtic cirrus emission,
all the detailed numbers shown above are applicable for the regions
except low-Galactic latitude zone.
}
\item We conclude that the FIR color-color diagrams are a powerful tool to 
separate stars from galaxies in AKARI FIS All-Sky Survey, and further, similar 
large FIR catalogs. 
With the presented simple method we can select more 
than 95~\% of galaxies, rejecting 80~\% of stars.
A more detailed methodology will be a subject of future works.
\end{enumerate}

\begin{acknowledgements}
First, we thank the anonymous referee for her/his careful reading and comments of the manuscript
which improved the clarity of the paper very much.
This work is based on observations with AKARI, a JAXA project with the participation of ESA. 
This research has made use of the NASA/IPAC Extragalactic Database (NED) which is operated 
by the Jet Propulsion Laboratory, California Institute of Technology, under contract with the 
National Aeronautics and Space Administration, and the SIMBAD database, operated at CDS, 
Strasbourg, France. 
We deeply thank Issei Yamamura for his huge effort of checking the reliability of our stellar 
identifications and their classification.
AP was financed by the research grant of the Polish Ministry of Science 
PBZ/MNiSW/07/2006/34A. 
TTT has been supported by Program for Improvement of Research
Environment for Young Researchers from Special Coordination Funds for
Promoting Science and Technology, and the Grant-in-Aid for the Scientific
Research Fund (20740105) commissioned by the Ministry of Education, Culture,
Sports, Science and Technology (MEXT) of Japan. 
TTT has been partially supported from the Grand-in-Aid for the Global COE Program 
``Quest for Fundamental Principles in the Universe: from Particles to the Solar System 
and the Cosmos'' from the MEXT.
\end{acknowledgements}

\begin{center}
\begin{longtable} {llcccc}
\caption{Consistency of star and galaxy identifications between different color-color diagrams}
\label{tab:consistency} \\
\hline
\hline
Diagram 1 & Diagram 2 & Galaxies & Stars & Galaxies & Stars \\
 & & in the galaxy cloud & in the star cloud & in the star cloud & in the galaxy cloud \\
 & & [\%] & [\%] & [\%] & [\%] \\
\hline
\endfirsthead
\hline
\hline
Diagram 1 & Diagram 2 & Galaxies & Stars & Galaxies & Stars \\
 & & in the galaxy cloud & in the star cloud & in the star cloud & in the galaxy cloud \\
 & & [\%] & [\%] & [\%] & [\%] \\
\hline
\endhead
\hline \multicolumn{6}{r}{{Continued on next page}} \\ \hline \hline
\endfoot
\endlastfoot
 $S_{65}/S_{90}$ vs. $S_{65}/S_{140}$ & $S_{65}/S_{90}$ vs. $S_{65}/S_{160}$ & 97.23 & 82.17 & 28.75 & 49.45 \\ 
 $S_{65}/S_{90}$ vs. $S_{65}/S_{140}$ & $S_{65}/S_{90}$ vs. $S_{90}/S_{140}$ & 99.76 & 99.15 & 94.01 & 95.58 \\ 
 $S_{65}/S_{90}$ vs. $S_{65}/S_{140}$ & $S_{65}/S_{90}$ vs. $S_{90}/S_{160}$ & 96.99 & 85.23 & 28.41 & 51.53 \\ 
 $S_{65}/S_{90}$ vs. $S_{65}/S_{140}$ & $S_{65}/S_{90}$ vs. $S_{140}/S_{160}$ & 97.14 & 63.13 & 26.79 & 39.39 \\ 
 $S_{65}/S_{90}$ vs. $S_{65}/S_{140}$ & $S_{65}/S_{140}$ vs. $S_{65}/S_{160}$ & 98.86 & 95.89 & 69.68 & 78.95 \\ 
 $S_{65}/S_{90}$ vs. $S_{65}/S_{140}$ & $S_{65}/S_{140}$ vs. $S_{90}/S_{140}$ & 99.30 & 96.81 & 78.36 & 86.57 \\ 
 $S_{65}/S_{90}$ vs. $S_{65}/S_{140}$ & $S_{65}/S_{140}$ vs. $S_{90}/S_{160}$ & 98.57 & 95.62 & 66.48 & 75.00 \\ 
 $S_{65}/S_{90}$ vs. $S_{65}/S_{140}$ & $S_{65}/S_{140}$ vs. $S_{140}/S_{160}$ & 97.05 & 93.40 & 41.75 & 63.55 \\ 
 $S_{65}/S_{90}$ vs. $S_{65}/S_{140}$ & $S_{65}/S_{160}$ vs. $S_{90}/S_{140}$ & 98.48 & 95.37 & 61.30 & 76.52 \\ 
 $S_{65}/S_{90}$ vs. $S_{65}/S_{140}$ & $S_{65}/S_{160}$ vs. $S_{90}/S_{160}$ & 97.02 & 76.23 & 18.54 & 44.76 \\ 
 $S_{65}/S_{90}$ vs. $S_{65}/S_{140}$ & $S_{65}/S_{160}$ vs. $S_{140}/S_{160}$ & 99.00 & 95.50 & 70.67 & 78.33 \\ 
 $S_{65}/S_{90}$ vs. $S_{65}/S_{140}$ & $S_{90}/S_{140}$ vs. $S_{90}/S_{160}$ & 98.37 & 93.75 & 60.12 & 70.49 \\ 
 $S_{65}/S_{90}$ vs. $S_{65}/S_{140}$ & $S_{90}/S_{140}$ vs. $S_{140}/S_{160}$ & 98.66 & 94.30 & 66.87 & 72.27 \\ 
 $S_{65}/S_{90}$ vs. $S_{65}/S_{140}$ & $S_{90}/S_{160}$ vs. $S_{140}/S_{160}$ & 98.43 & 94.66 & 63.04 & 73.50 \\ 
 $S_{65}/S_{90}$ vs. $S_{65}/S_{160}$ & $S_{65}/S_{90}$ vs. $S_{90}/S_{140}$ & 97.22 & 82.53 & 32.54 & 48.59 \\ 
 $S_{65}/S_{90}$ vs. $S_{65}/S_{160}$ & $S_{65}/S_{90}$ vs. $S_{90}/S_{160}$ & 99.57 & 95.97 & 89.97 & 91.63 \\ 
 $S_{65}/S_{90}$ vs. $S_{65}/S_{160}$ & $S_{65}/S_{90}$ vs. $S_{140}/S_{160}$ & 98.33 & 77.30 & 57.85 & 74.39 \\ 
 $S_{65}/S_{90}$ vs. $S_{65}/S_{160}$ & $S_{65}/S_{140}$ vs. $S_{65}/S_{160}$ & 97.96 & 84.23 & 46.50 & 53.93 \\ 
 $S_{65}/S_{90}$ vs. $S_{65}/S_{160}$ & $S_{65}/S_{140}$ vs. $S_{90}/S_{140}$ & 97.44 & 80.40 & 22.06 & 50.51 \\ 
 $S_{65}/S_{90}$ vs. $S_{65}/S_{160}$ & $S_{65}/S_{140}$ vs. $S_{90}/S_{160}$ & 97.50 & 83.02 & 41.93 & 46.43 \\ 
 $S_{65}/S_{90}$ vs. $S_{65}/S_{160}$ & $S_{65}/S_{140}$ vs. $S_{140}/S_{160}$ & 97.71 & 84.25 & 55.29 & 51.46 \\ 
 $S_{65}/S_{90}$ vs. $S_{65}/S_{160}$ & $S_{65}/S_{160}$ vs. $S_{90}/S_{140}$ & 97.16 & 80.92 & 28.75 & 44.69 \\ 
 $S_{65}/S_{90}$ vs. $S_{65}/S_{160}$ & $S_{65}/S_{160}$ vs. $S_{90}/S_{160}$ & 98.64 & 82.55 & 63.40 & 72.99 \\ 
 $S_{65}/S_{90}$ vs. $S_{65}/S_{160}$ & $S_{65}/S_{160}$ vs. $S_{140}/S_{160}$ & 97.95 & 84.05 & 41.11 & 55.43 \\ 
 $S_{65}/S_{90}$ vs. $S_{65}/S_{160}$ & $S_{90}/S_{140}$ vs. $S_{90}/S_{160}$ & 96.78 & 79.30 & 22.35 & 43.01 \\ 
 $S_{65}/S_{90}$ vs. $S_{65}/S_{160}$ & $S_{90}/S_{140}$ vs. $S_{140}/S_{160}$ & 96.64 & 79.22 & 17.86 & 41.53 \\ 
 $S_{65}/S_{90}$ vs. $S_{65}/S_{160}$ & $S_{90}/S_{160}$ vs. $S_{140}/S_{160}$ & 96.59 & 79.69 & 20.96 & 41.99 \\ 
 $S_{65}/S_{90}$ vs. $S_{90}/S_{140}$ & $S_{65}/S_{90}$ vs. $S_{90}/S_{160}$ & 97.01 & 85.93 & 32.51 & 51.90 \\ 
 $S_{65}/S_{90}$ vs. $S_{90}/S_{140}$ & $S_{65}/S_{90}$ vs. $S_{140}/S_{160}$ & 97.14 & 63.33 & 30.68 & 37.84 \\ 
 $S_{65}/S_{90}$ vs. $S_{90}/S_{140}$ & $S_{65}/S_{140}$ vs. $S_{65}/S_{160}$ & 98.83 & 95.76 & 70.73 & 77.06 \\ 
 $S_{65}/S_{90}$ vs. $S_{90}/S_{140}$ & $S_{65}/S_{140}$ vs. $S_{90}/S_{140}$ & 99.06 & 95.96 & 72.73 & 82.17 \\ 
 $S_{65}/S_{90}$ vs. $S_{90}/S_{140}$ & $S_{65}/S_{140}$ vs. $S_{90}/S_{160}$ & 98.54 & 95.83 & 67.57 & 74.75 \\ 
 $S_{65}/S_{90}$ vs. $S_{90}/S_{140}$ & $S_{65}/S_{140}$ vs. $S_{140}/S_{160}$ & 97.07 & 93.62 & 44.65 & 62.75 \\ 
 $S_{65}/S_{90}$ vs. $S_{90}/S_{140}$ & $S_{65}/S_{160}$ vs. $S_{90}/S_{140}$ & 98.33 & 94.90 & 59.82 & 72.73 \\ 
 $S_{65}/S_{90}$ vs. $S_{90}/S_{140}$ & $S_{65}/S_{160}$ vs. $S_{90}/S_{160}$ & 96.96 & 76.27 & 21.88 & 42.93 \\ 
 $S_{65}/S_{90}$ vs. $S_{90}/S_{140}$ & $S_{65}/S_{160}$ vs. $S_{140}/S_{160}$ & 98.92 & 95.03 & 70.43 & 74.78 \\ 
 $S_{65}/S_{90}$ vs. $S_{90}/S_{140}$ & $S_{90}/S_{140}$ vs. $S_{90}/S_{160}$ & 98.19 & 93.29 & 58.19 & 66.67 \\ 
 $S_{65}/S_{90}$ vs. $S_{90}/S_{140}$ & $S_{90}/S_{140}$ vs. $S_{140}/S_{160}$ & 98.44 & 93.84 & 63.43 & 68.42 \\ 
 $S_{65}/S_{90}$ vs. $S_{90}/S_{140}$ & $S_{90}/S_{160}$ vs. $S_{140}/S_{160}$ & 98.28 & 94.20 & 61.58 & 69.64 \\ 
 $S_{65}/S_{90}$ vs. $S_{90}/S_{160}$ & $S_{65}/S_{90}$ vs. $S_{140}/S_{160}$ & 97.90 & 73.52 & 50.86 & 66.67 \\ 
 $S_{65}/S_{90}$ vs. $S_{90}/S_{160}$ & $S_{65}/S_{140}$ vs. $S_{65}/S_{160}$ & 97.79 & 87.20 & 46.61 & 56.60 \\ 
 $S_{65}/S_{90}$ vs. $S_{90}/S_{160}$ & $S_{65}/S_{140}$ vs. $S_{90}/S_{140}$ & 97.17 & 83.24 & 21.55 & 51.40 \\ 
 $S_{65}/S_{90}$ vs. $S_{90}/S_{160}$ & $S_{65}/S_{140}$ vs. $S_{90}/S_{160}$ & 97.43 & 86.70 & 44.44 & 51.01 \\ 
 $S_{65}/S_{90}$ vs. $S_{90}/S_{160}$ & $S_{65}/S_{140}$ vs. $S_{140}/S_{160}$ & 97.89 & 87.91 & 61.22 & 56.58 \\ 
 $S_{65}/S_{90}$ vs. $S_{90}/S_{160}$ & $S_{65}/S_{160}$ vs. $S_{90}/S_{140}$ & 97.05 & 84.01 & 31.25 & 46.25 \\ 
 $S_{65}/S_{90}$ vs. $S_{90}/S_{160}$ & $S_{65}/S_{160}$ vs. $S_{90}/S_{160}$ & 98.70 & 85.33 & 67.67 & 74.51 \\ 
 $S_{65}/S_{90}$ vs. $S_{90}/S_{160}$ & $S_{65}/S_{160}$ vs. $S_{140}/S_{160}$ & 97.76 & 87.05 & 41.03 & 58.18 \\ 
 $S_{65}/S_{90}$ vs. $S_{90}/S_{160}$ & $S_{90}/S_{140}$ vs. $S_{90}/S_{160}$ & 96.61 & 82.49 & 24.11 & 44.31 \\ 
 $S_{65}/S_{90}$ vs. $S_{90}/S_{160}$ & $S_{90}/S_{140}$ vs. $S_{140}/S_{160}$ & 96.42 & 82.40 & 18.84 & 42.68 \\ 
 $S_{65}/S_{90}$ vs. $S_{90}/S_{160}$ & $S_{90}/S_{160}$ vs. $S_{140}/S_{160}$ & 96.42 & 82.84 & 22.75 & 43.21 \\ 
 $S_{65}/S_{90}$ vs. $S_{140}/S_{160}$ & $S_{65}/S_{140}$ vs. $S_{65}/S_{160}$ & 97.29 & 63.01 & 29.21 & 37.69 \\ 
 $S_{65}/S_{90}$ vs. $S_{140}/S_{160}$ & $S_{65}/S_{140}$ vs. $S_{90}/S_{140}$ & 97.21 & 60.77 & 15.38 & 41.43 \\ 
 $S_{65}/S_{90}$ vs. $S_{140}/S_{160}$ & $S_{65}/S_{140}$ vs. $S_{90}/S_{160}$ & 96.73 & 61.61 & 24.29 & 31.20 \\ 
 $S_{65}/S_{90}$ vs. $S_{140}/S_{160}$ & $S_{65}/S_{140}$ vs. $S_{140}/S_{160}$ & 96.27 & 62.92 & 27.34 & 34.78 \\ 
 $S_{65}/S_{90}$ vs. $S_{140}/S_{160}$ & $S_{65}/S_{160}$ vs. $S_{90}/S_{140}$ & 96.52 & 59.50 & 12.80 & 32.18 \\ 
 $S_{65}/S_{90}$ vs. $S_{140}/S_{160}$ & $S_{65}/S_{160}$ vs. $S_{90}/S_{160}$ & 97.32 & 57.31 & 28.01 & 58.99 \\ 
 $S_{65}/S_{90}$ vs. $S_{140}/S_{160}$ & $S_{65}/S_{160}$ vs $S_{140}/S_{160}$ & 97.46 & 62.96 & 27.08 & 39.85 \\ 
 $S_{65}/S_{90}$ vs. $S_{140}/S_{160}$ & $S_{90}/S_{140}$ vs. $S_{90}/S_{160}$ & 96.23 & 57.67 & 9.38 & 32.09 \\ 
 $S_{65}/S_{90}$ vs. $S_{140}/S_{160}$ & $S_{90}/S_{140}$ vs $S_{140}/S_{160}$ & 96.21 & 57.74 & 7.72 & 30.94 \\ 
 $S_{65}/S_{90}$ vs. $S_{140}/S_{160}$ & $S_{90}/S_{160}$ vs. $S_{140}/S_{160}$ & 96.07 & 57.93 & 9.04 & 30.42 \\ 
 $S_{65}/S_{140}$ vs. $S_{65}/S_{160}$ & $S_{65}/S_{140}$ vs. $S_{90}/S_{140}$ & 98.96 & 95.07 & 67.18 & 78.46 \\ 
 $S_{65}/S_{140}$ vs. $S_{65}/S_{160}$ & $S_{65}/S_{140}$ vs. $S_{90}/S_{160}$ & 99.45 & 98.33 & 86.88 & 90.00 \\ 
 $S_{65}/S_{140}$ vs. $S_{65}/S_{160}$ & $S_{65}/S_{140}$ vs. $S_{140}/S_{160}$ & 98.13 & 96.81 & 62.56 & 81.55 \\ 
 $S_{65}/S_{140}$ vs. $S_{65}/S_{160}$ & $S_{65}/S_{160}$ vs. $S_{90}/S_{140}$ & 98.99 & 97.10 & 73.82 & 84.68 \\ 
 $S_{65}/S_{140}$ vs. $S_{65}/S_{160}$ & $S_{65}/S_{160}$ vs. $S_{90}/S_{160}$ & 97.99 & 80.08 & 43.92 & 52.43 \\ 
 $S_{65}/S_{140}$ vs. $S_{65}/S_{160}$ & $S_{65}/S_{160}$ vs. $S_{140}/S_{160}$ & 99.67 & 98.97 & 90.25 & 94.83 \\ 
 $S_{65}/S_{140}$ vs. $S_{65}/S_{160}$ & $S_{90}/S_{140}$ vs. $S_{90}/S_{160}$ & 98.61 & 95.52 & 65.45 & 77.97 \\ 
 $S_{65}/S_{140}$ vs. $S_{65}/S_{160}$ & $S_{90}/S_{140}$ vs. $S_{140}/S_{160}$ & 98.52 & 95.71 & 62.58 & 78.26 \\ 
 $S_{65}/S_{140}$ vs. $S_{65}/S_{160}$ & $S_{90}/S_{160}$ vs. $S_{140}/S_{160}$ & 98.52 & 96.07 & 64.72 & 79.65 \\ 
 $S_{65}/S_{140}$ vs. $S_{90}/S_{140}$ & $S_{65}/S_{140}$ vs. $S_{90}/S_{160}$ & 98.70 & 94.46 & 64.45 & 73.33 \\ 
 $S_{65}/S_{140}$ vs. $S_{90}/S_{140}$ & $S_{65}/S_{140}$ vs. $S_{140}/S_{160}$ & 97.21 & 91.83 & 37.36 & 61.79 \\ 
 $S_{65}/S_{140}$ vs. $S_{90}/S_{140}$ & $S_{65}/S_{160}$ vs. $S_{90}/S_{140}$ & 98.92 & 95.59 & 67.64 & 80.92 \\ 
 $S_{65}/S_{140}$ vs. $S_{90}/S_{140}$ & $S_{65}/S_{160}$ vs. $S_{90}/S_{160}$ & 97.39 & 76.27 & 14.96 & 50.44 \\ 
 $S_{65}/S_{140}$ vs. $S_{90}/S_{140}$ & $S_{65}/S_{160}$ vs. $S_{140}/S_{160}$ & 99.22 & 95.37 & 72.34 & 80.88 \\ 
 $S_{65}/S_{140}$ vs. $S_{90}/S_{140}$ & $S_{90}/S_{140}$ vs. $S_{90}/S_{160}$ & 98.81 & 94.64 & 65.97 & 78.26 \\ 
 $S_{65}/S_{140}$ vs. $S_{90}/S_{140}$ & $S_{90}/S_{140}$ vs. $S_{140}/S_{160}$ & 99.01 & 95.56 & 71.13 & 81.48 \\ 
 $S_{65}/S_{140}$ vs. $S_{90}/S_{140}$ & $S_{90}/S_{160}$ vs. $S_{140}/S_{160}$ & 98.80 & 95.22 & 67.11 & 79.70 \\ 
 $S_{65}/S_{140}$ vs. $S_{90}/S_{160}$ & $S_{65}/S_{140}$ vs. $S_{140}/S_{160}$ & 98.31 & 97.19 & 69.21 & 81.72 \\ 
 $S_{65}/S_{140}$ vs. $S_{90}/S_{160}$ & $S_{65}/S_{160}$ vs. $S_{90}/S_{140}$ & 99.34 & 97.82 & 84.83 & 87.13 \\ 
 $S_{65}/S_{140}$ vs. $S_{90}/S_{160}$ & $S_{65}/S_{160}$ vs. $S_{90}/S_{160}$ & 97.89 & 78.88 & 48.36 & 45.92 \\ 
 $S_{65}/S_{140}$ vs. $S_{90}/S_{160}$ & $S_{65}/S_{160}$ vs. $S_{140}/S_{160}$ & 99.20 & 97.30 & 79.11 & 84.91 \\ 
 $S_{65}/S_{140}$ vs. $S_{90}/S_{160}$ & $S_{90}/S_{140}$ vs. $S_{90}/S_{160}$ & 98.86 & 96.27 & 74.80 & 79.63 \\ 
 $S_{65}/S_{140}$ vs. $S_{90}/S_{160}$ & $S_{90}/S_{140}$ vs. $S_{140}/S_{160}$ & 98.57 & 96.46 & 67.95 & 80.00 \\ 
 $S_{65}/S_{140}$ vs. $S_{90}/S_{160}$ & $S_{90}/S_{160}$ vs. $S_{140}/S_{160}$ & 98.77 & 96.81 & 73.82 & 81.55 \\ 
 $S_{65}/S_{140}$ vs. $S_{140}/S_{160}$ & $S_{65}/S_{160}$ vs. $S_{90}/S_{140}$ & 97.87 & 95.29 & 58.71 & 73.08 \\ 
 $S_{65}/S_{140}$ vs. $S_{140}/S_{160}$ & $S_{65}/S_{160}$ vs. $S_{90}/S_{160}$ & 98.48 & 79.36 & 68.84 & 48.24 \\ 
 $S_{65}/S_{140}$ vs. $S_{140}/S_{160}$ & $S_{65}/S_{160}$ vs. $S_{140}/S_{160}$ & 97.98 & 96.43 & 56.46 & 80.73 \\ 
 $S_{65}/S_{140}$ vs. $S_{140}/S_{160}$ & $S_{90}/S_{140}$ vs. $S_{90}/S_{160}$ & 97.31 & 94.04 & 49.54 & 68.47 \\ 
 $S_{65}/S_{140}$ vs. $S_{140}/S_{160}$ & $S_{90}/S_{140}$ vs. $S_{140}/S_{160}$ & 96.94 & 93.90 & 42.06 & 66.67 \\ 
 $S_{65}/S_{140}$ vs. $S_{140}/S_{160}$ & $S_{90}/S_{160}$ vs. $S_{140}/S_{160}$ & 97.17 & 94.26 & 48.54 & 67.92 \\ 
 $S_{65}/S_{160}$ vs. $S_{90}/S_{140}$ & $S_{65}/S_{160}$ vs. $S_{90}/S_{160}$ & 97.70 & 79.02 & 38.83 & 50.24 \\ 
 $S_{65}/S_{160}$ vs. $S_{90}/S_{140}$ & $S_{65}/S_{160}$ vs. $S_{140}/S_{160}$ & 98.96 & 97.07 & 70.34 & 85.47 \\ 
 $S_{65}/S_{160}$ vs. $S_{90}/S_{140}$ & $S_{90}/S_{140}$ vs. $S_{90}/S_{160}$ & 99.43 & 98.45 & 86.30 & 92.44 \\ 
 $S_{65}/S_{160}$ vs. $S_{90}/S_{140}$ & $S_{90}/S_{140}$ vs. $S_{140}/S_{160}$ & 99.09 & 98.28 & 77.88 & 91.38 \\ 
 $S_{65}/S_{160}$ vs. $S_{90}/S_{140}$ & $S_{90}/S_{160}$ vs. $S_{140}/S_{160}$ & 99.29 & 98.97 & 83.71 & 94.74 \\ 
 $S_{65}/S_{160}$ vs. $S_{90}/S_{160}$ & $S_{65}/S_{160}$ vs. $S_{140}/S_{160}$ & 98.01 & 80.25 & 38.66 & 54.72 \\ 
 $S_{65}/S_{160}$ vs. $S_{90}/S_{160}$ & $S_{90}/S_{140}$ vs. $S_{90}/S_{160}$ & 97.28 & 79.34 & 30.43 & 53.27 \\ 
 $S_{65}/S_{160}$ vs. $S_{90}/S_{160}$ & $S_{90}/S_{140}$ vs. $S_{140}/S_{160}$ & 96.89 & 78.03 & 19.50 & 49.29 \\ 
 $S_{65}/S_{160}$ vs. $S_{90}/S_{160}$ & $S_{90}/S_{160}$ vs. $S_{140}/S_{160}$ & 97.02 & 78.53 & 26.87 & 49.76 \\ 
 $S_{65}/S_{160}$ vs. $S_{140}/S_{160}$ & $S_{90}/S_{140}$ vs. $S_{90}/S_{160}$ & 98.65 & 95.47 & 63.37 & 79.03 \\ 
 $S_{65}/S_{160}$ vs $S_{140}/S_{160}$ & $S_{90}/S_{140}$ vs. $S_{140}/S_{160}$ & 98.63 & 95.67 & 62.21 & 79.34 \\ 
 $S_{65}/S_{160}$ vs. $S_{140}/S_{160}$ & $S_{90}/S_{160}$ vs. $S_{140}/S_{160}$ & 98.57 & 96.03 & 62.66 & 80.67 \\ 
 $S_{90}/S_{140}$ vs. $S_{90}/S_{160}$ & $S_{90}/S_{140}$ vs. $S_{140}/S_{160}$ & 99.51 & 98.78 & 88.64 & 94.31 \\ 
 $S_{90}/S_{140}$ vs. $S_{90}/S_{160}$ & $S_{90}/S_{160}$ vs. $S_{140}/S_{160}$ & 99.74 & 98.79 & 94.31 & 94.21 \\ 
 $S_{90}/S_{140}$ vs. $S_{140}/S_{160}$ & $S_{90}/S_{160}$ vs. $S_{140}/S_{160}$ & 99.67 & 99.31 & 92.60 & 96.61 \\ 
\hline
\end{longtable}
\end{center}

\clearpage

\appendix

\section{Other flux-color diagrams}

%\clearpage
\begin{figure}[thb]
	\centering
	\includegraphics[width=0.45\textwidth]{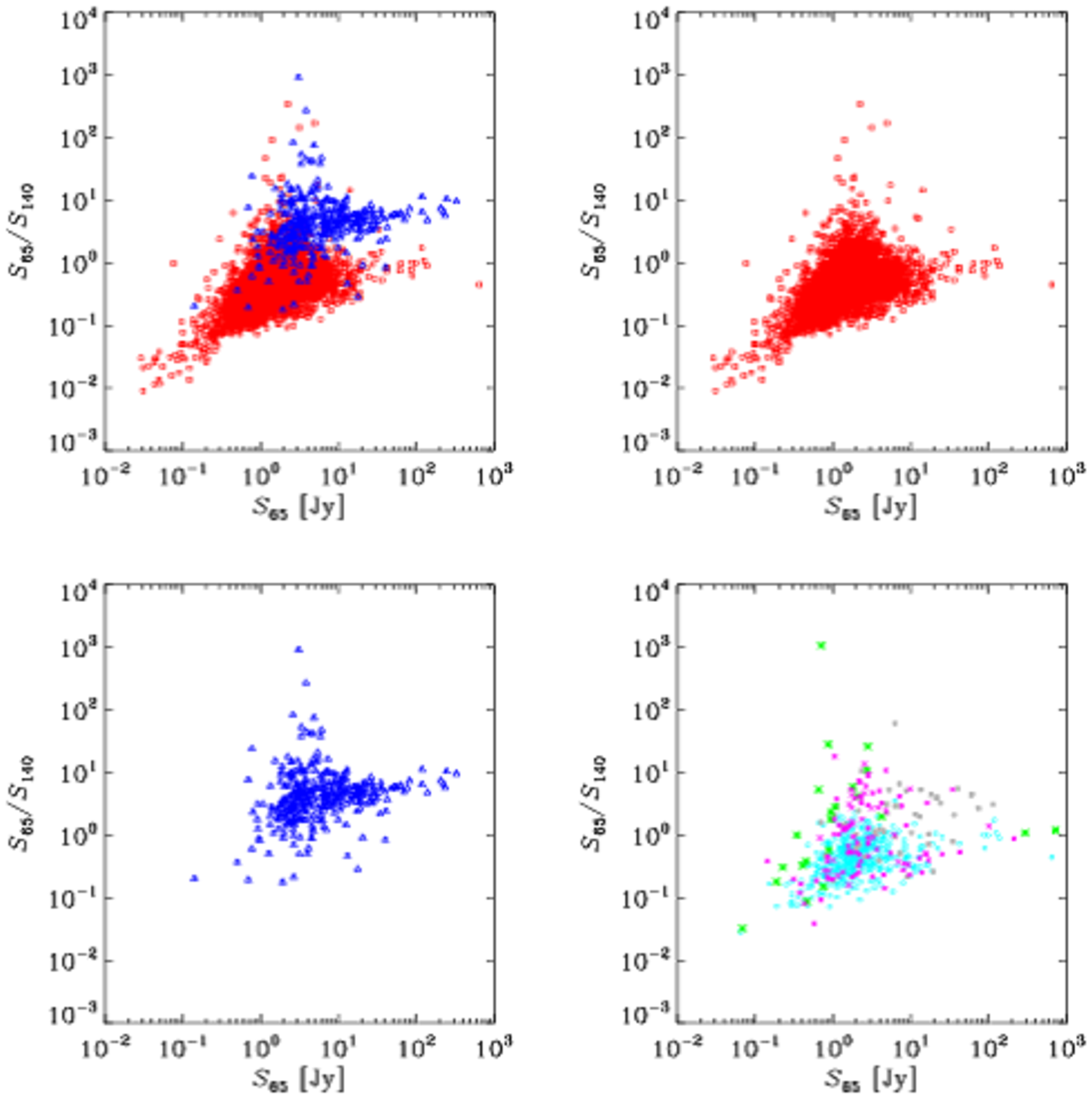}
	\includegraphics[width=0.45\textwidth]{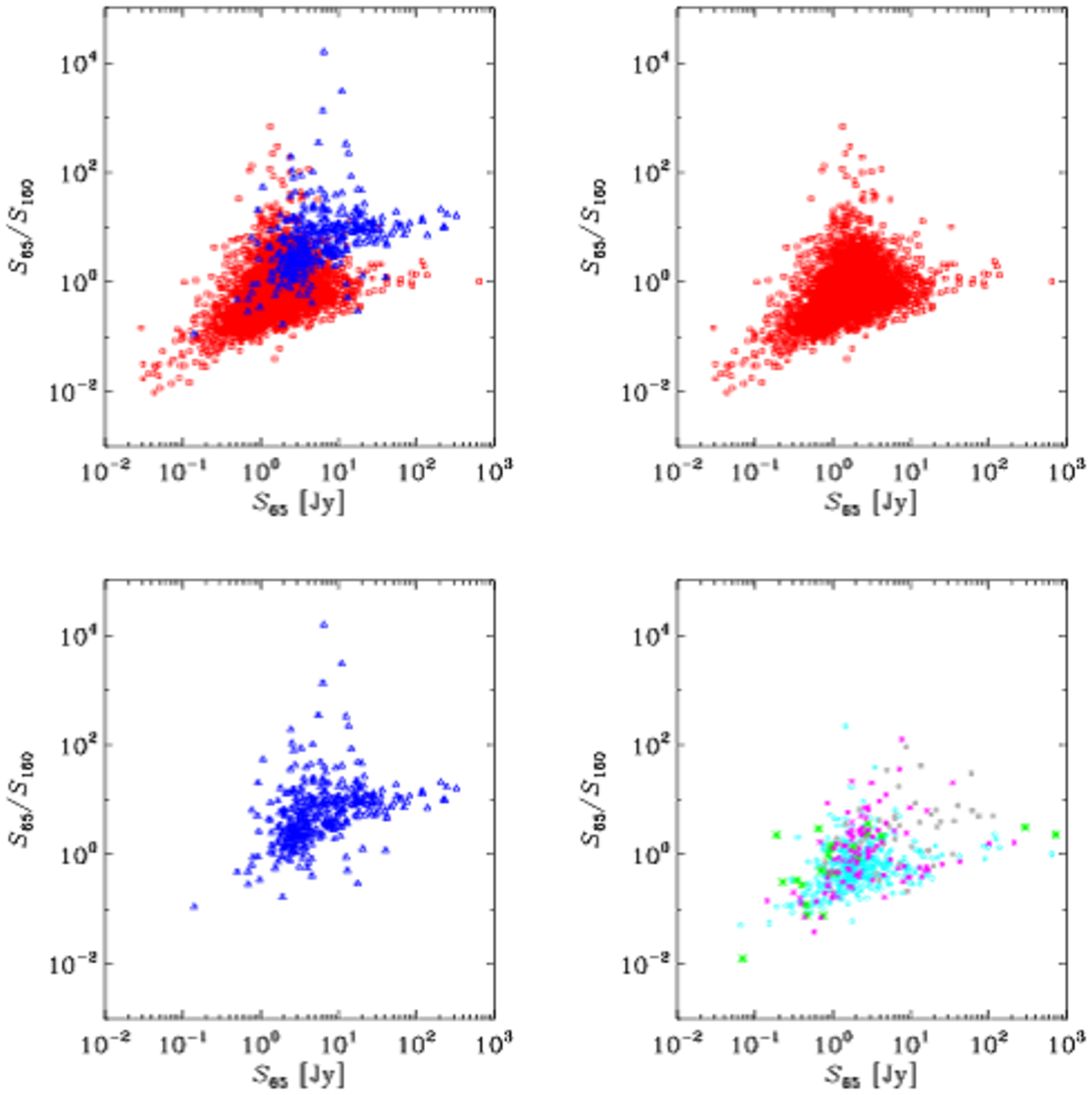}
	\caption{
	Same as Fig.~\ref{fig:fc1} but for 
	$S_{65}$--$S_{65}/S_{140}$ (four panels on the left side) and 
	for $S_{65}$--$S_{65}/S_{160}$ (four panels on the right side).
	}\label{fig:afc1}
\end{figure}

\begin{figure}[thb]
	\centering
	\includegraphics[width=0.45\textwidth]{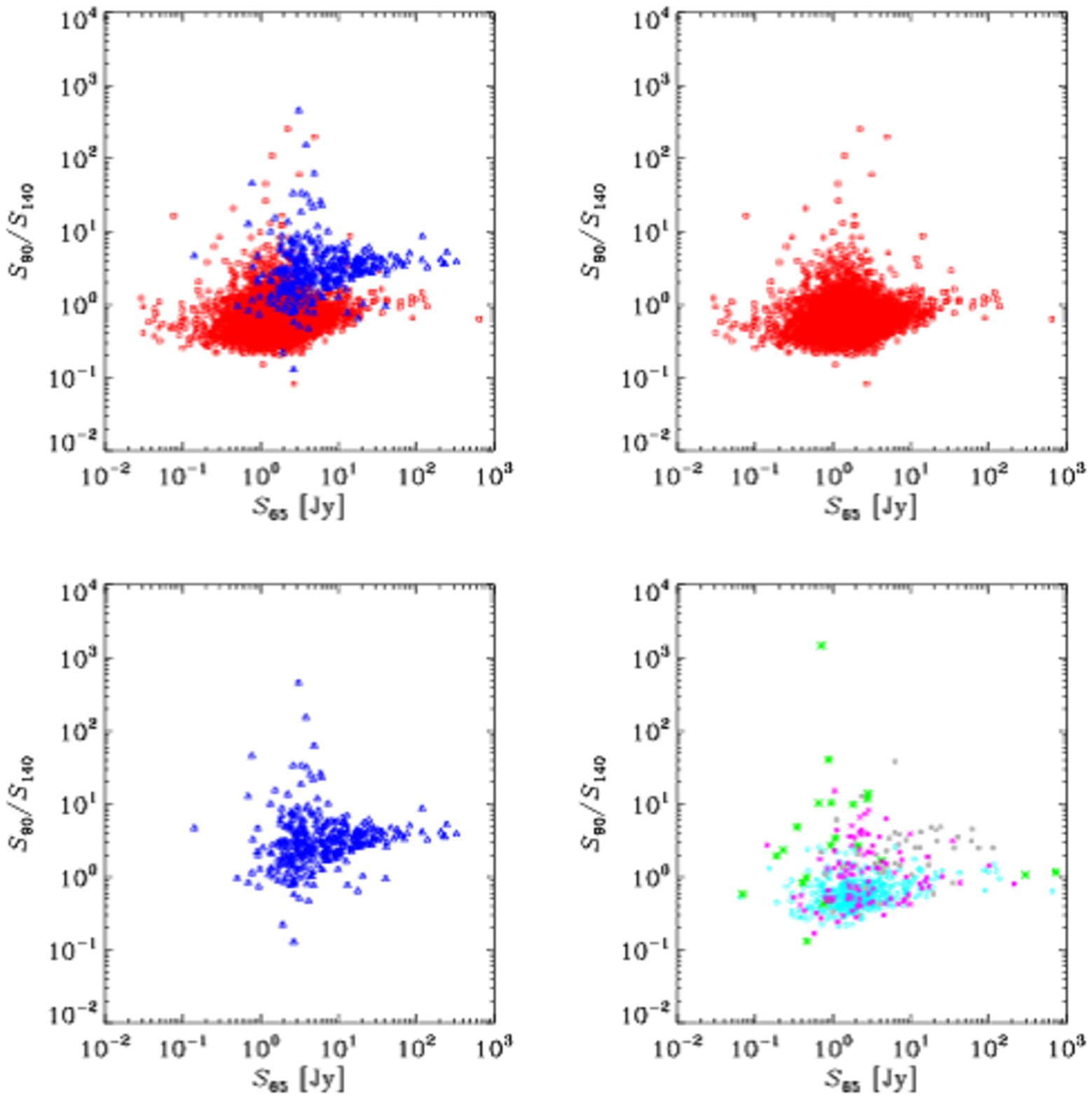}
	\includegraphics[width=0.45\textwidth]{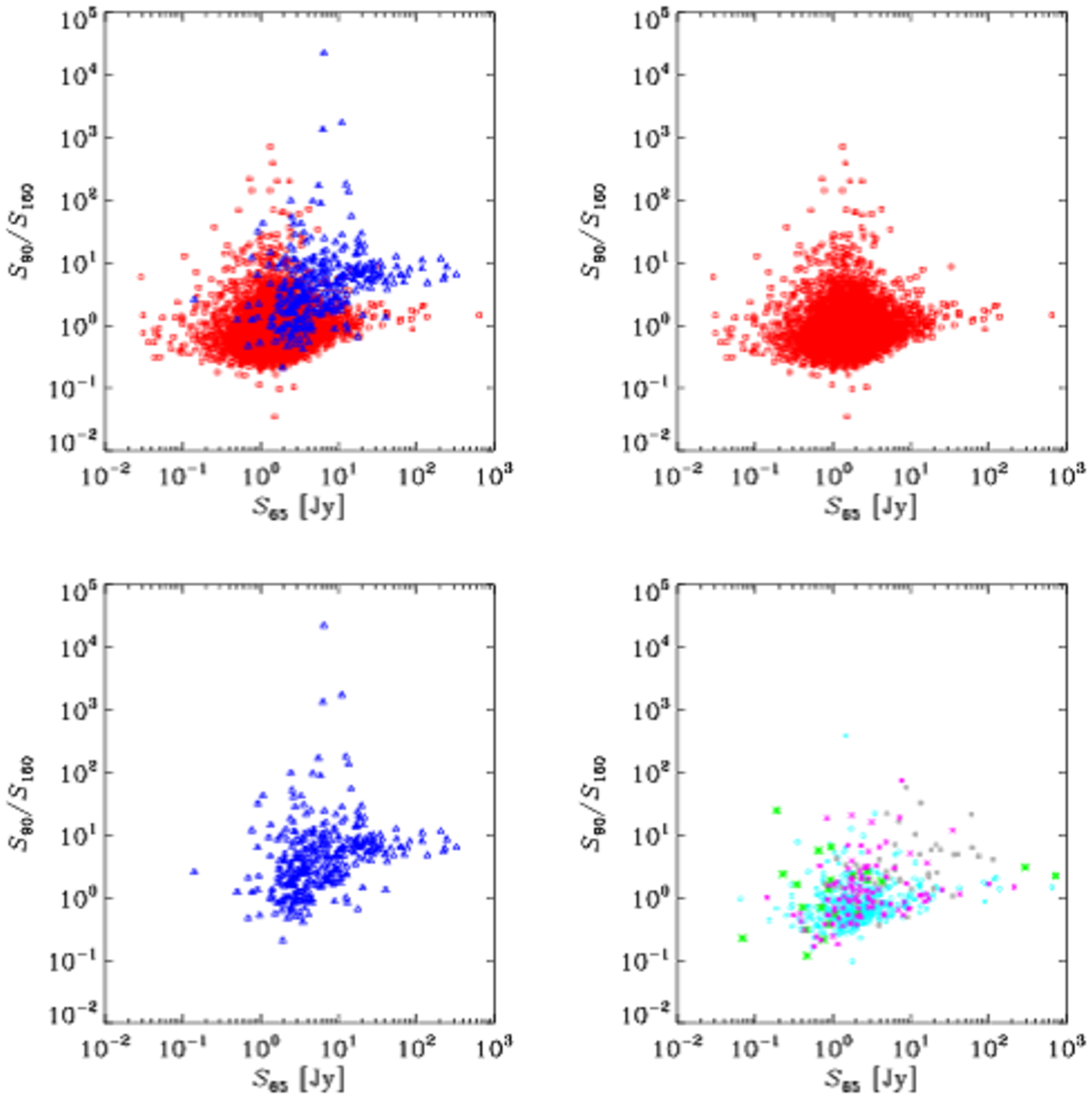}
	\caption{Same as Fig.~\ref{fig:fc1} but for 
	$S_{65}$--$S_{90}/S_{140}$ (four panels on the left side) and 
	$S_{65}$--$S_{90}/S_{160}$ (four panels on the right side).
	}\label{fig:afc2}
\end{figure}

\clearpage

%\onfig{1}{figure
\begin{figure}[thb]
	\centering
	\includegraphics[width=0.45\textwidth]{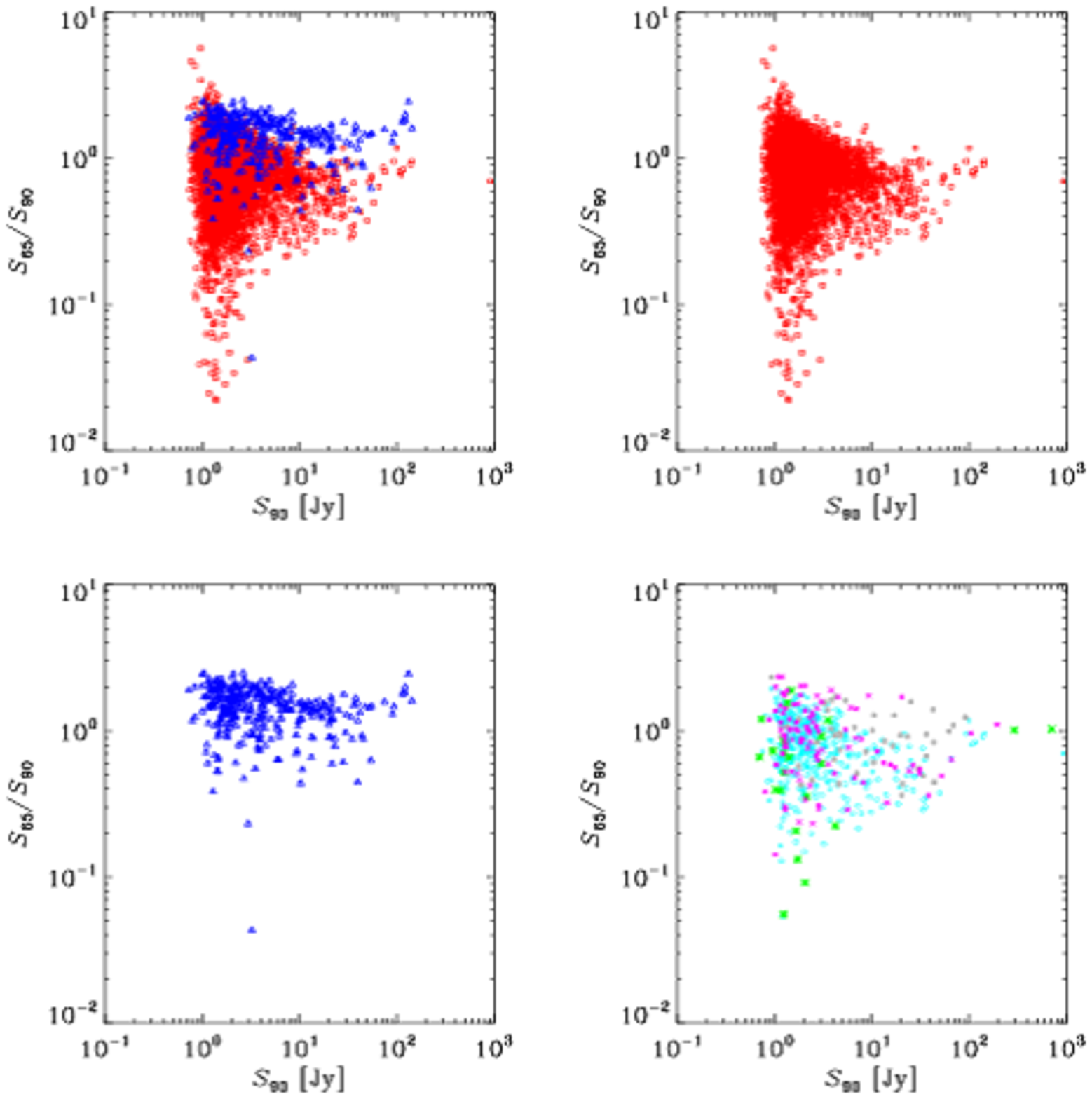}
	\includegraphics[width=0.45\textwidth]{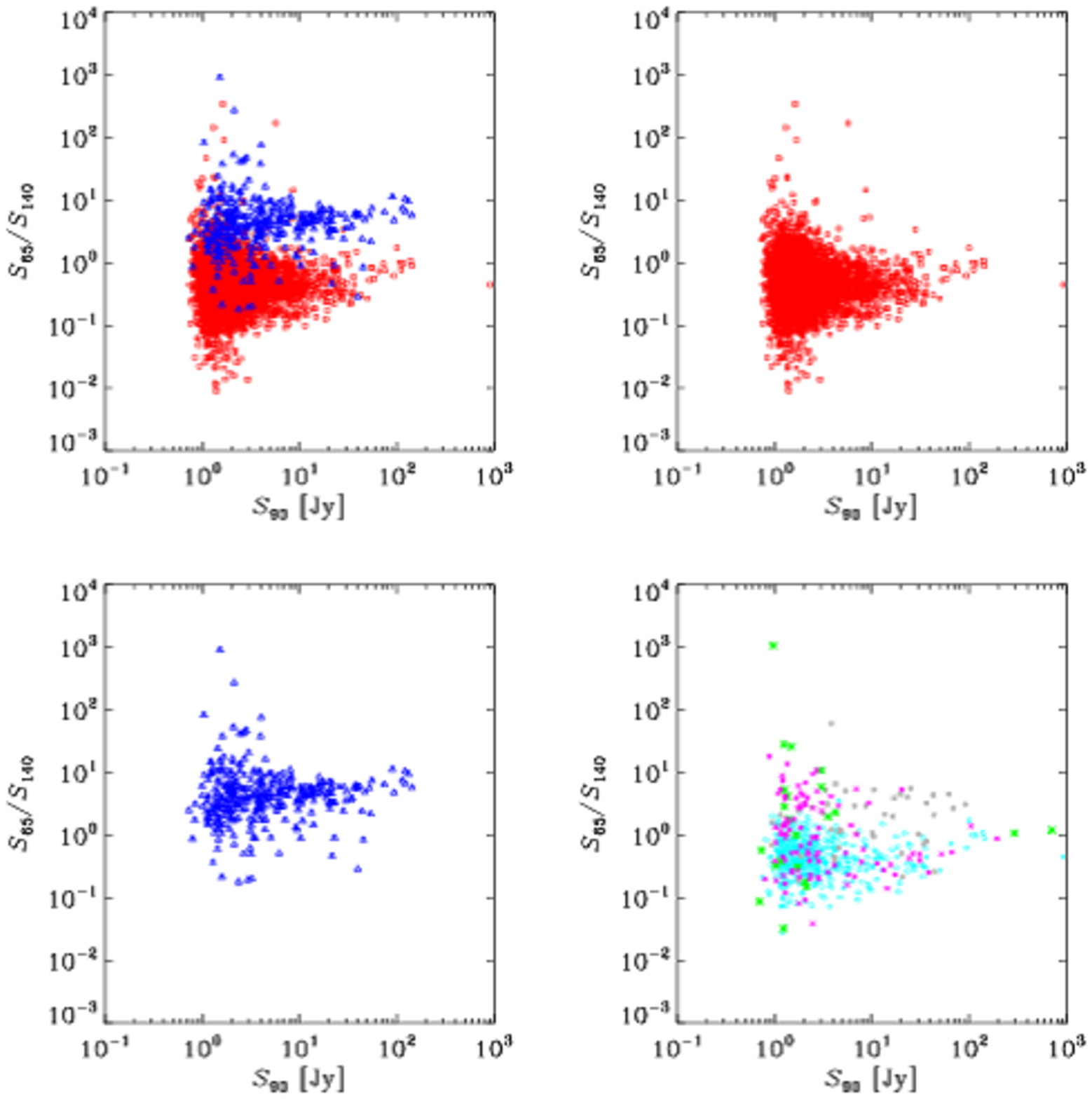}
	\caption{Same as Fig.~\ref{fig:fc1} but for 
	$S_{90}$--$S_{65}/S_{90}$ (four panels on the left side) and 
	$S_{90}$--$S_{65}/S_{140}$ (four panels on the right side). }  
	 \label{fig:afc3}
\end{figure}
%}

%\onfig{2}{
\begin{figure}[thb]
	\centering
	\includegraphics[width=0.45\textwidth]{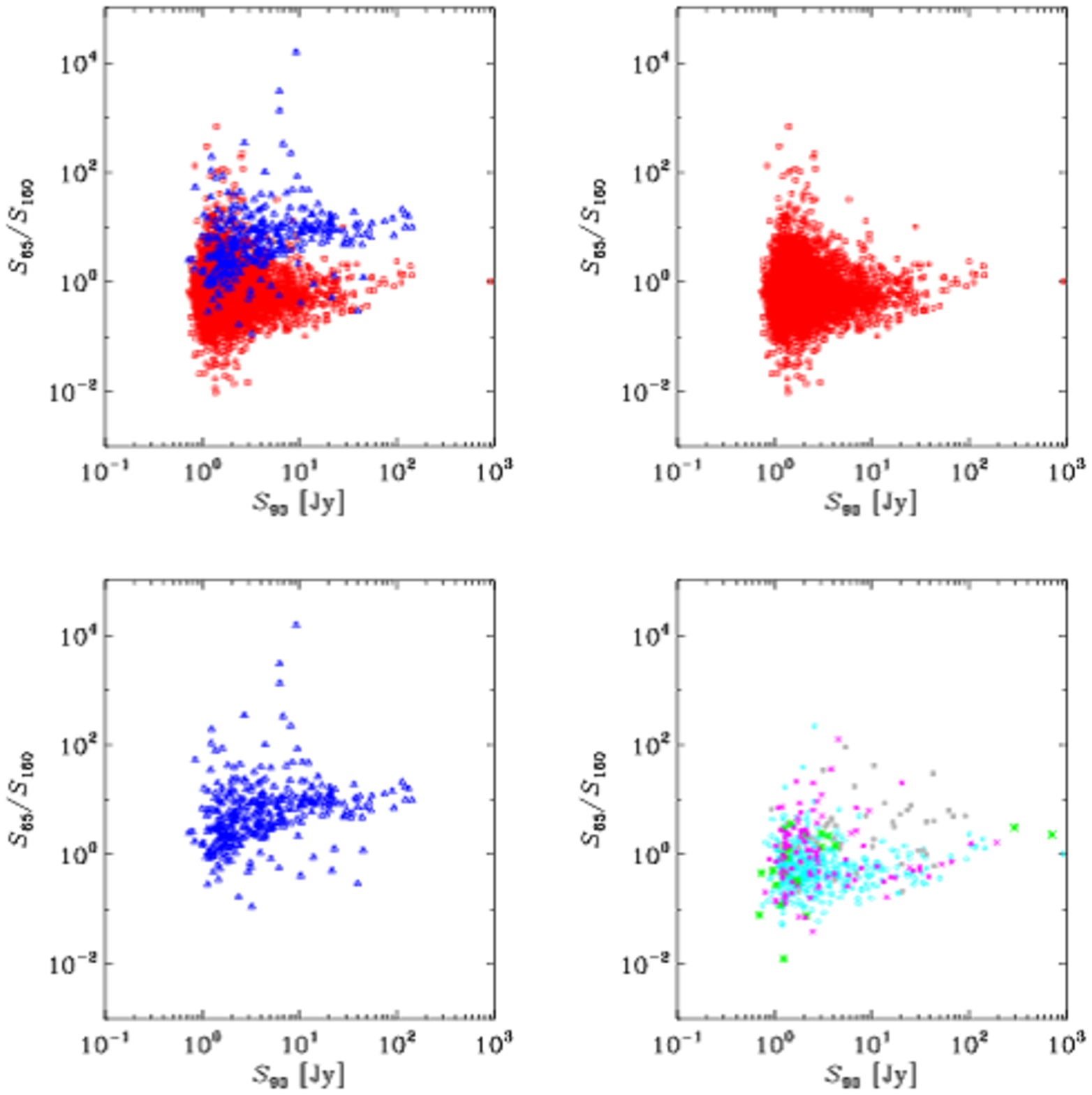}
	\includegraphics[width=0.45\textwidth]{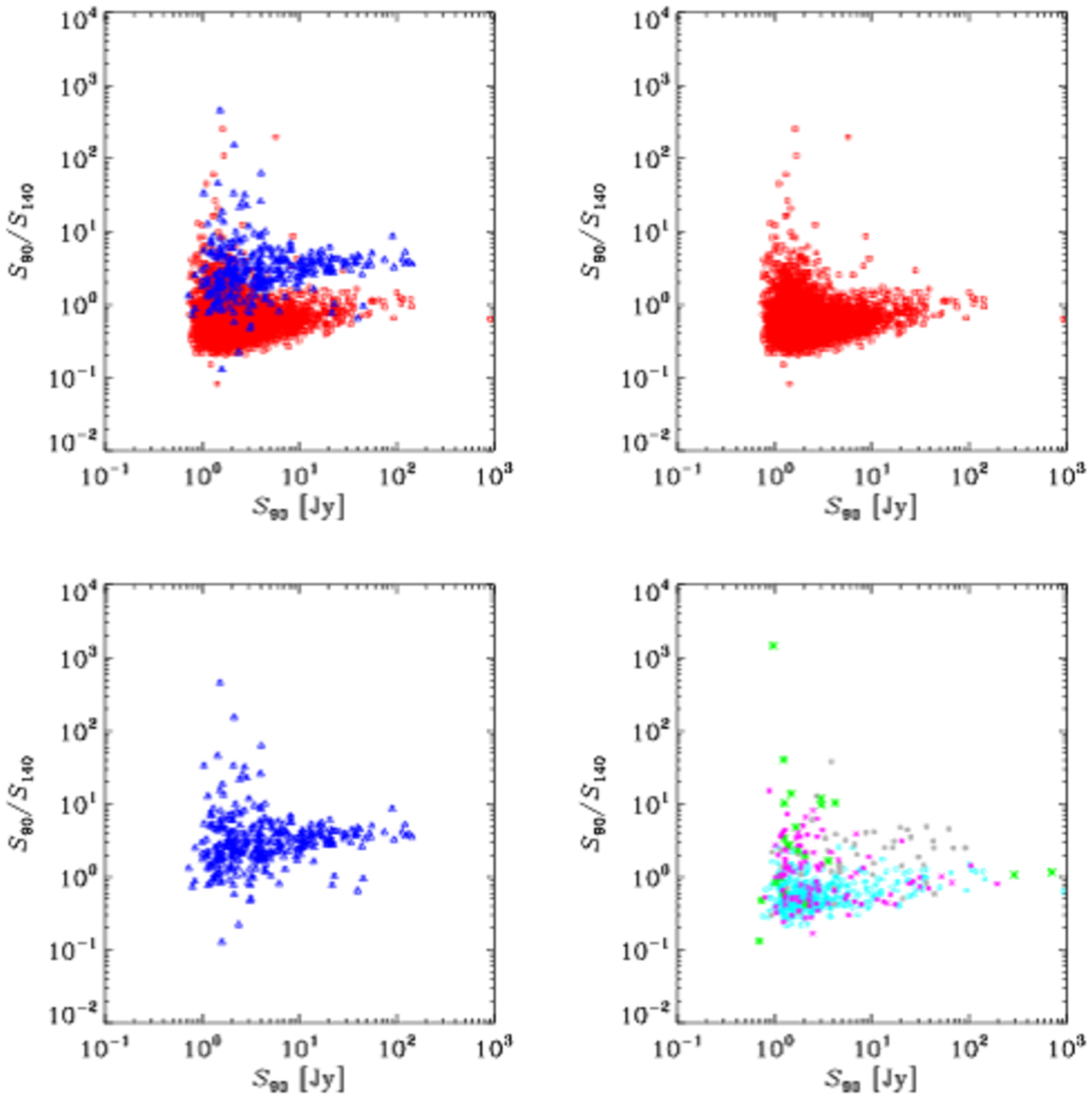}
	\caption{Same as Fig.~\ref{fig:fc1} but for 
	$S_{90}$--$S_{65}/S_{160}$ (four panels on the left side) and 
	$S_{90}$--$S_{90}/S_{140}$ (four panels on the right side). 
	}  
	\label{fig:afc4}
\end{figure}
%}

%\onfig{3}{
\begin{figure}[thb]
	\centering
	\includegraphics[width=0.45\textwidth]{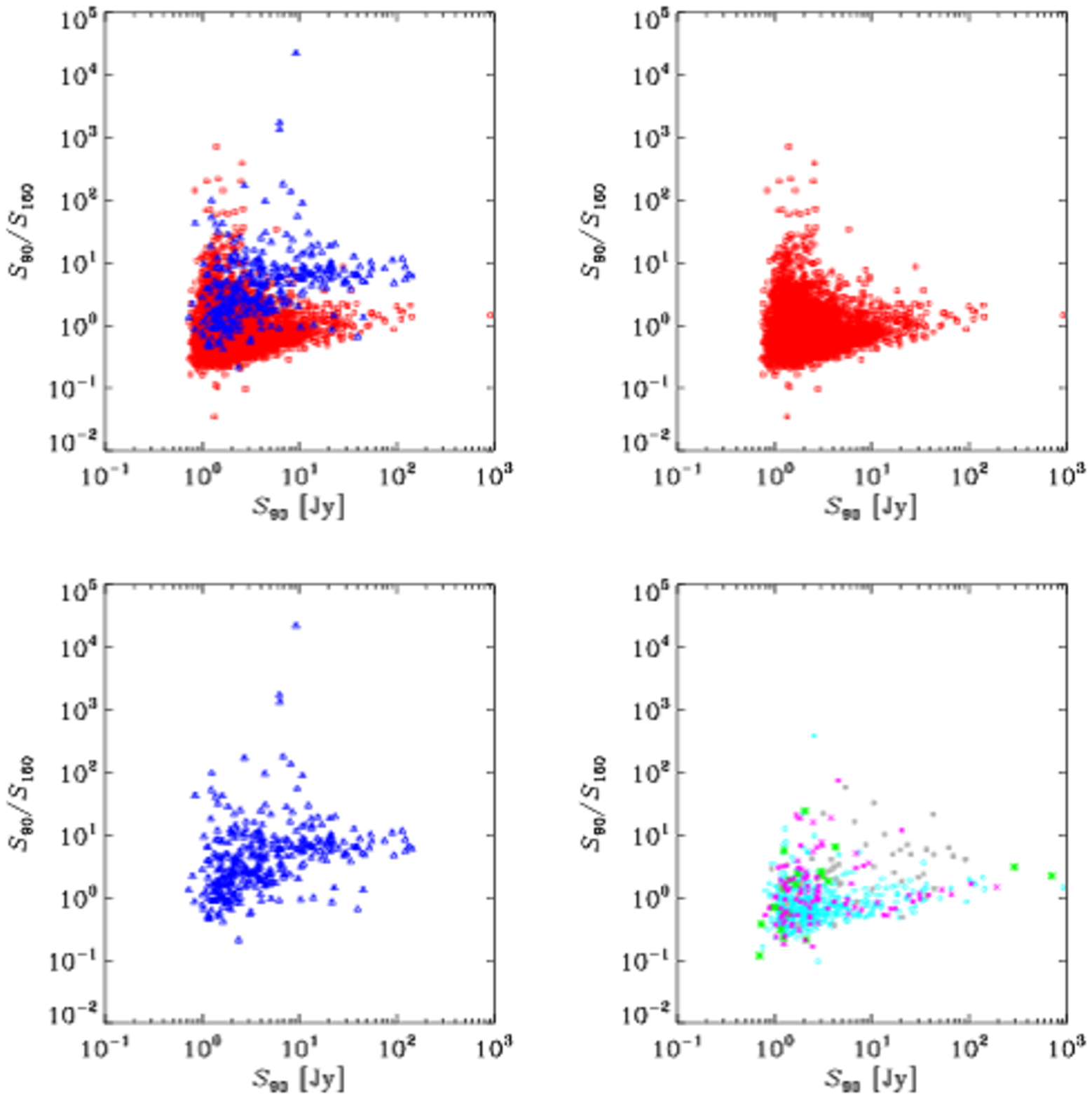}
	\includegraphics[width=0.45\textwidth]{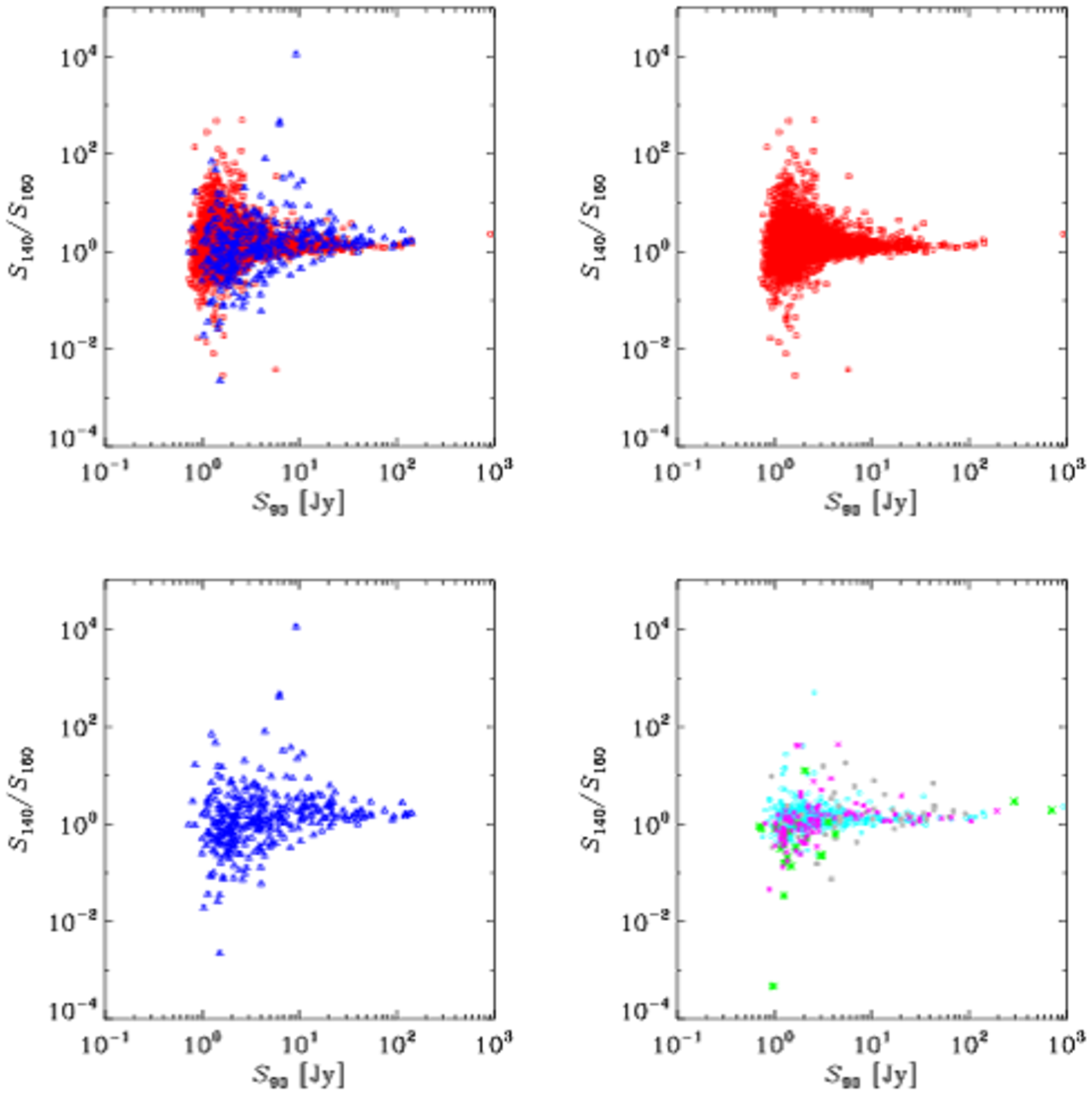}
	\caption{Same as Fig.~\ref{fig:fc1} but for 
	$S_{90}$--$S_{90}/S_{160}$ (four panels on the left side) and 
	$S_{90}$--$S_{140}/S_{160}$ (four panels on the right side).  
	}  
	\label{fig:afc5}
\end{figure}
%}

%\onfig{4}{
\begin{figure}[thb]
	\centering
	\includegraphics[width=0.45\textwidth]{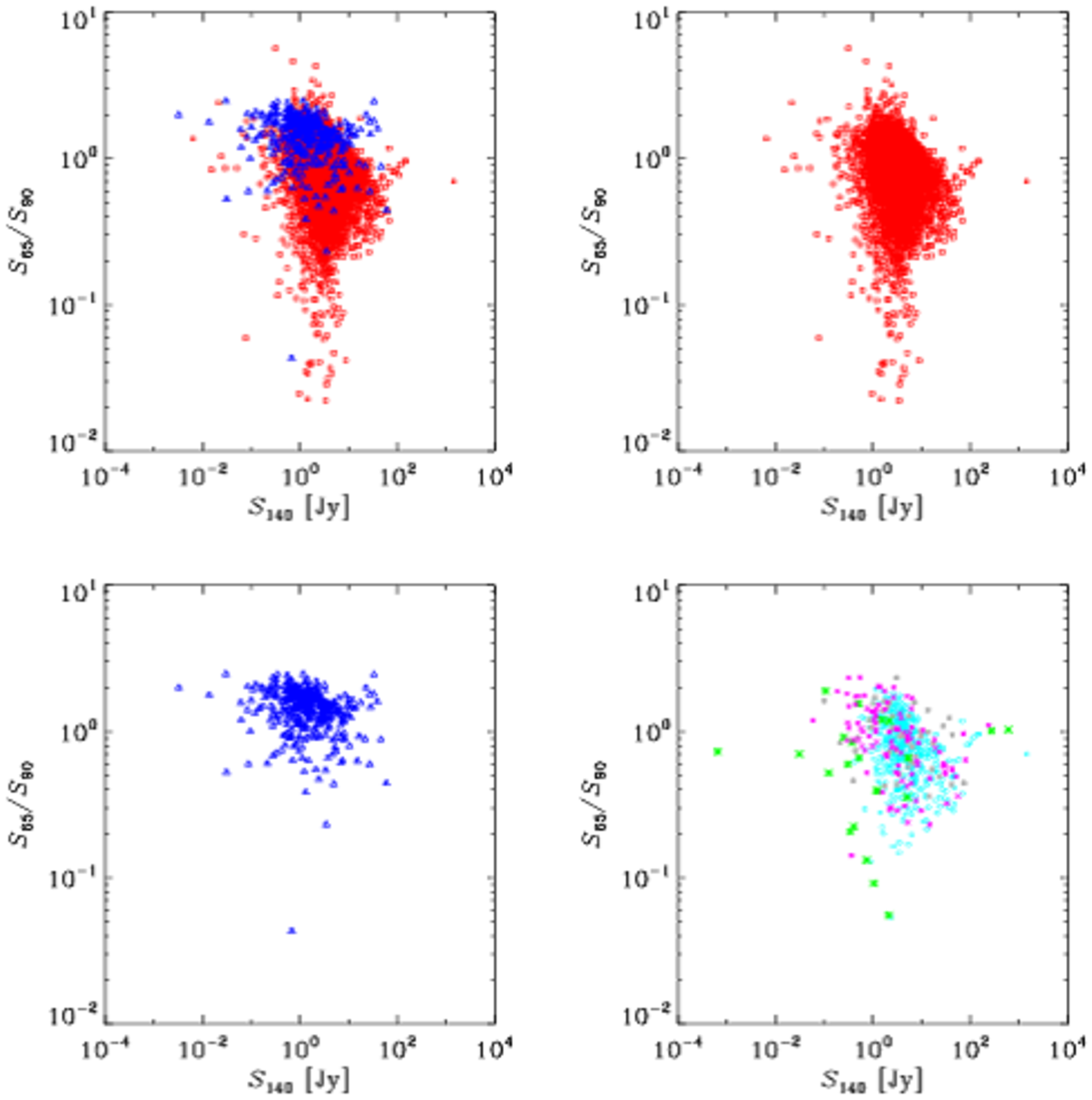}
	\includegraphics[width=0.45\textwidth]{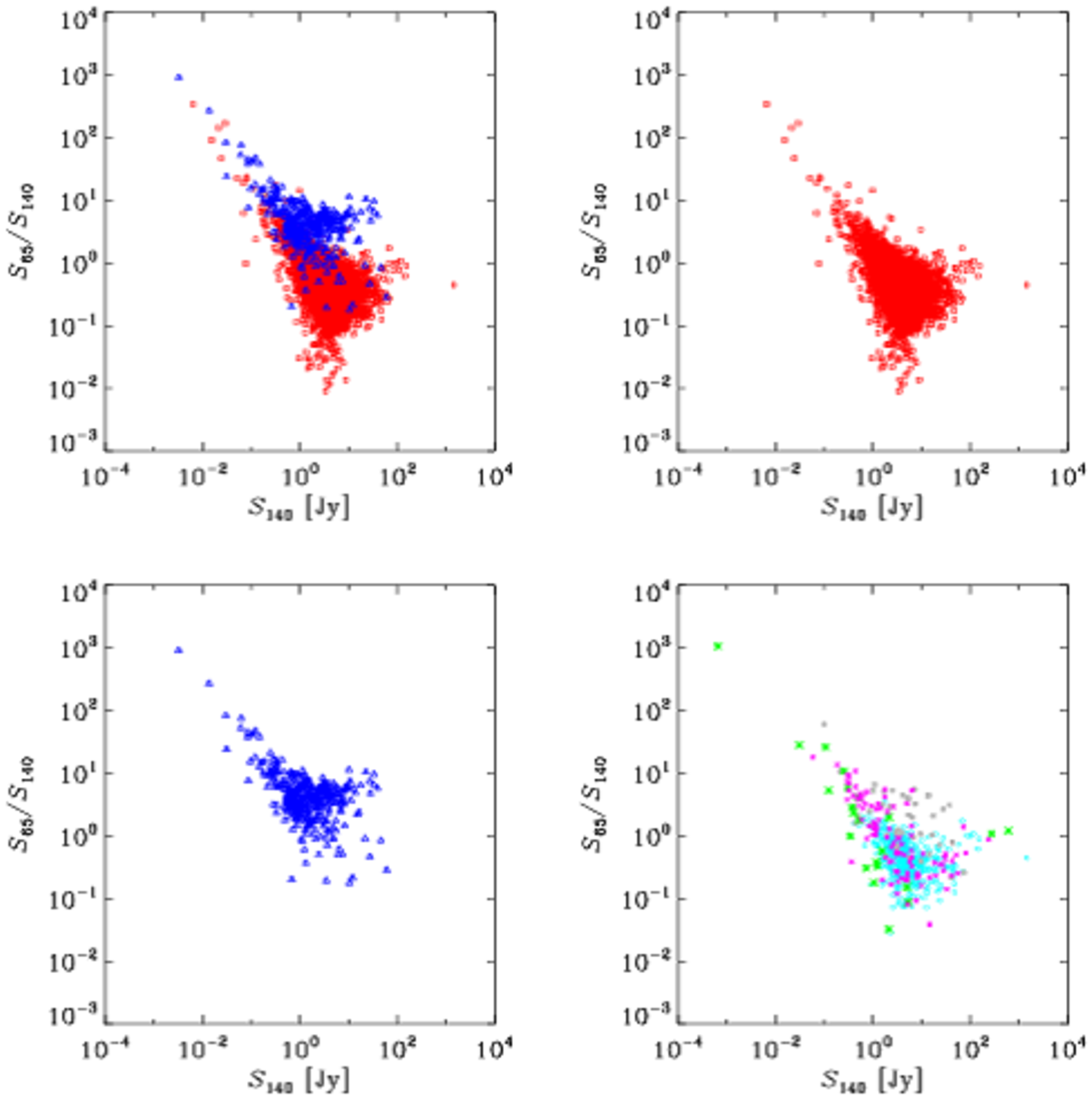}
	\caption{Same as Fig.~\ref{fig:fc1} but for 
	$S_{140}$--$S_{65}/S_{90}$ (four panels on the left side) and 
	$S_{140}$--$S_{65}/S_{140}$ (four panels on the right side). 
	}
	\label{fig:afc6}
\end{figure}
%}

%\onfig{5}{
\begin{figure}[thb]
	\centering
	\includegraphics[width=0.45\textwidth]{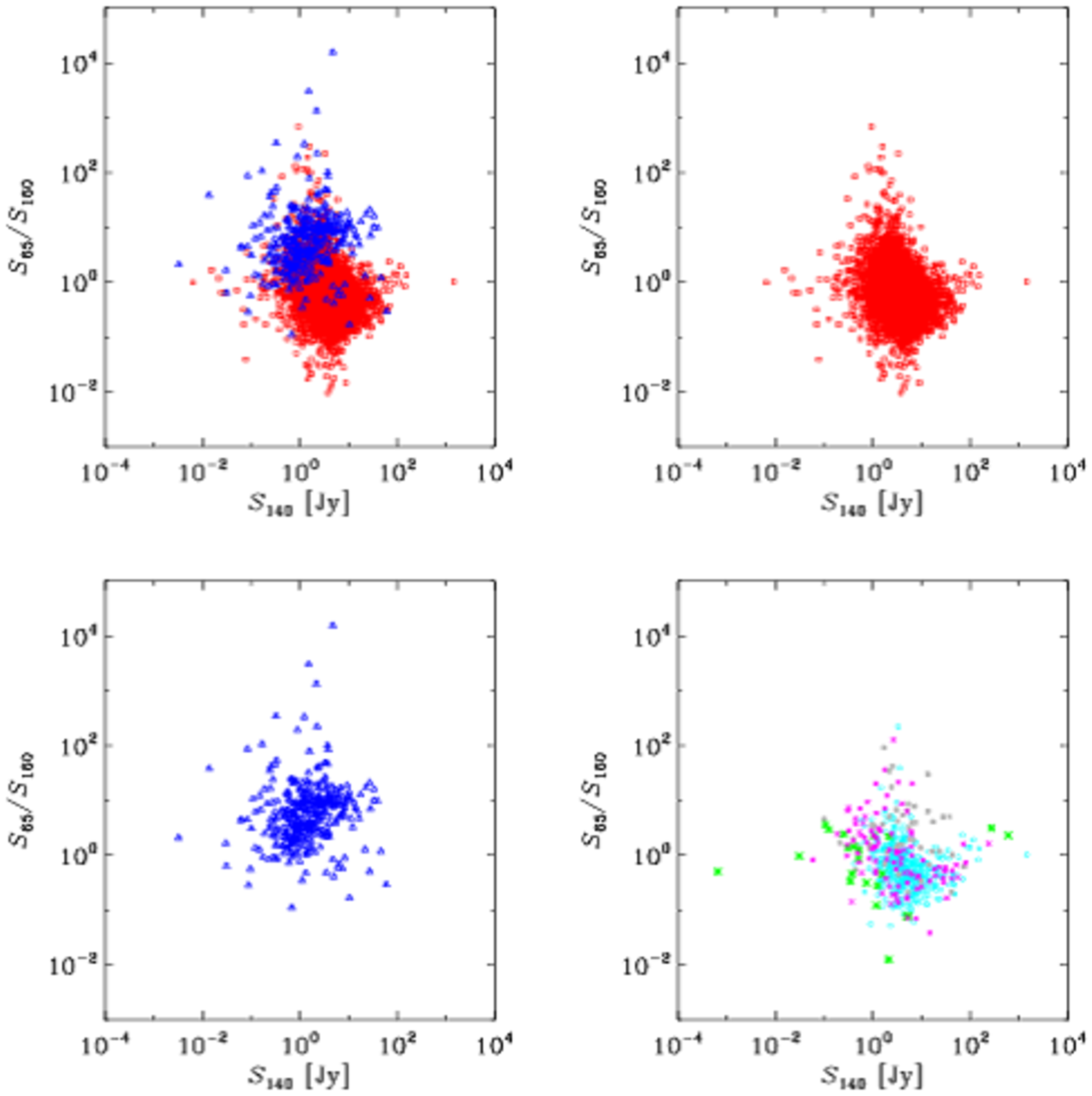}
	\includegraphics[width=0.45\textwidth]{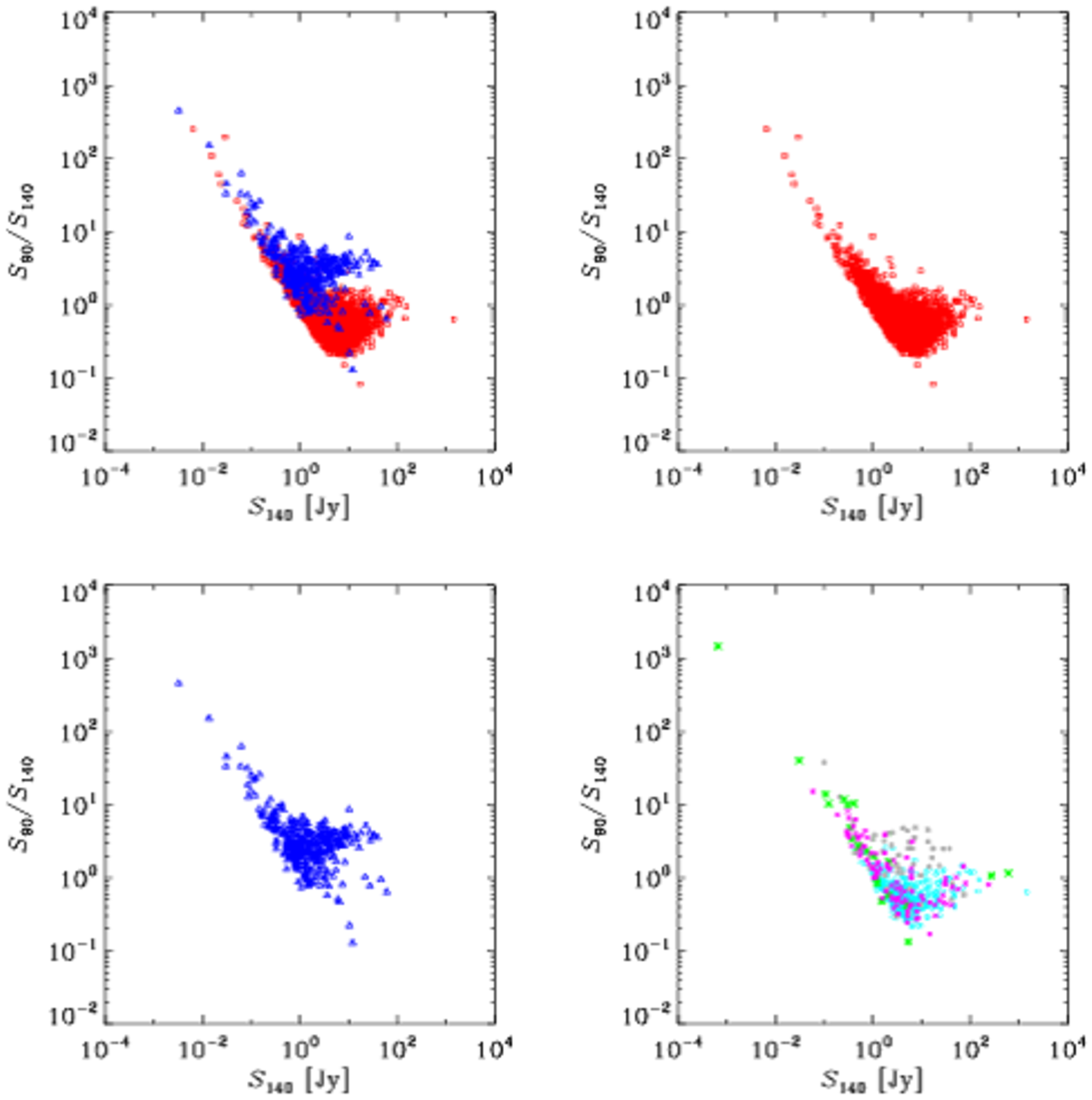}
	\caption{Same as Fig.~\ref{fig:fc1} but for 
	$S_{140}$--$S_{65}/S_{160}$ (four panels on the left side) and 
	$S_{140}$--$S_{90}/S_{140}$ (four panels on the right side).  }  
	 \label{fig:afc7}
\end{figure}
%}

%\onfig{6}{
\begin{figure}[thb]
	\centering
	\includegraphics[width=0.45\textwidth]{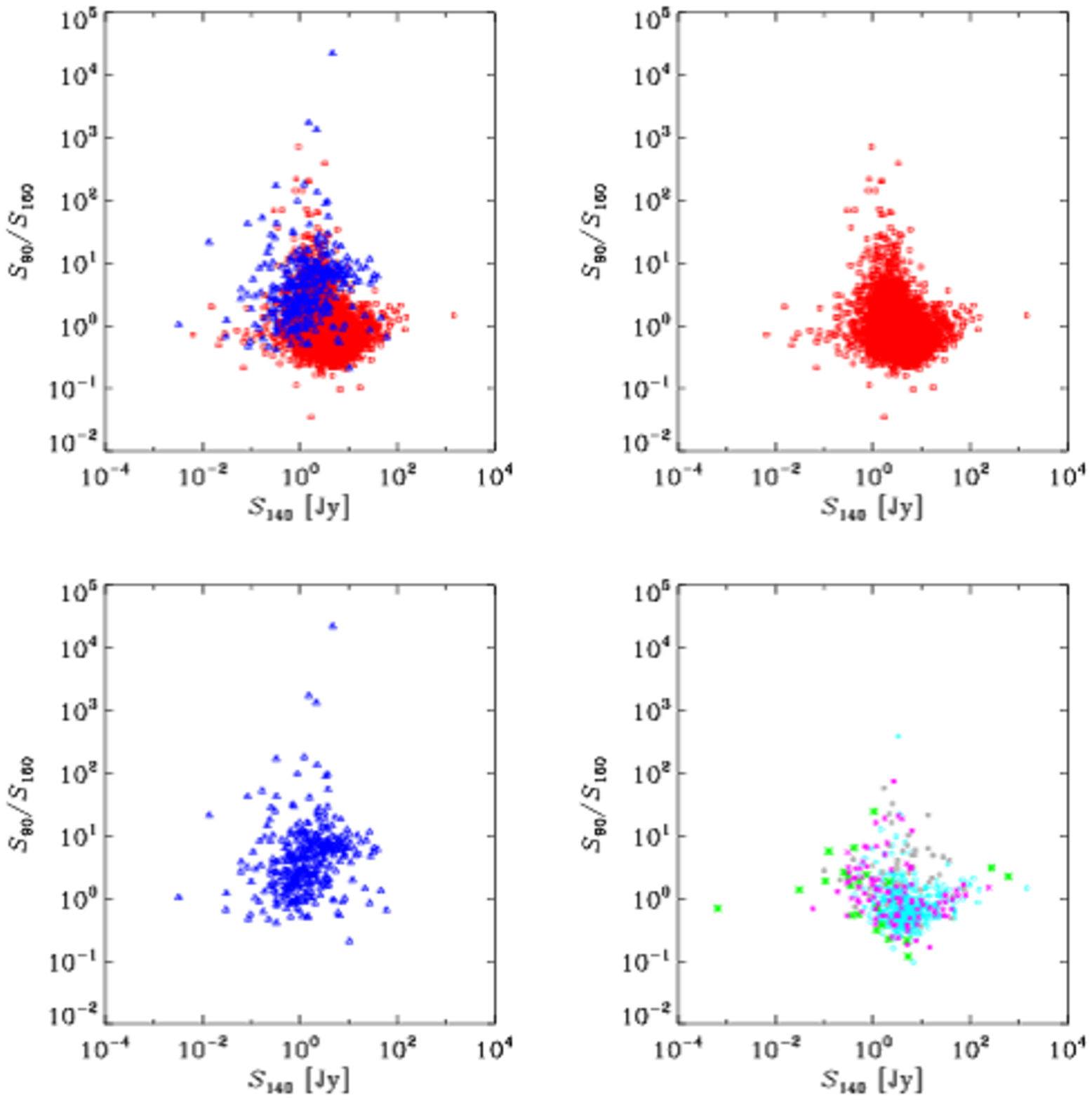}
	\includegraphics[width=0.45\textwidth]{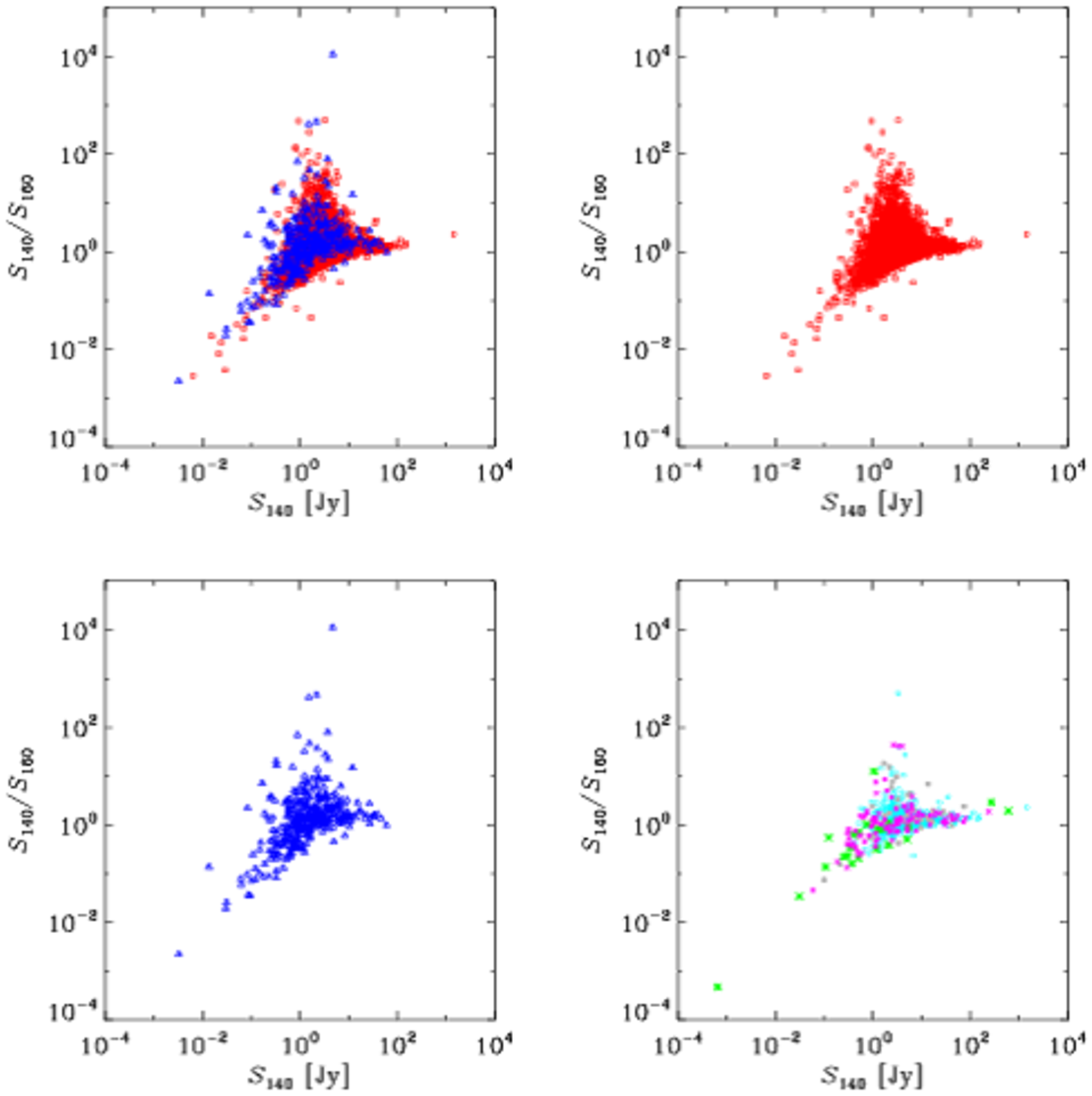}
	\caption{Same as Fig.~\ref{fig:fc1} but for 
	$S_{140}$--$S_{90}/S_{160}$ (four panels on the left side) and 
	$S_{140}$--$S_{140}/S_{160}$ (four panels on the right side). }  
	 \label{fig:afc8}
\end{figure}
%}

%\onfig{7}{
\begin{figure}[thb]
	\centering
	\includegraphics[width=0.45\textwidth]{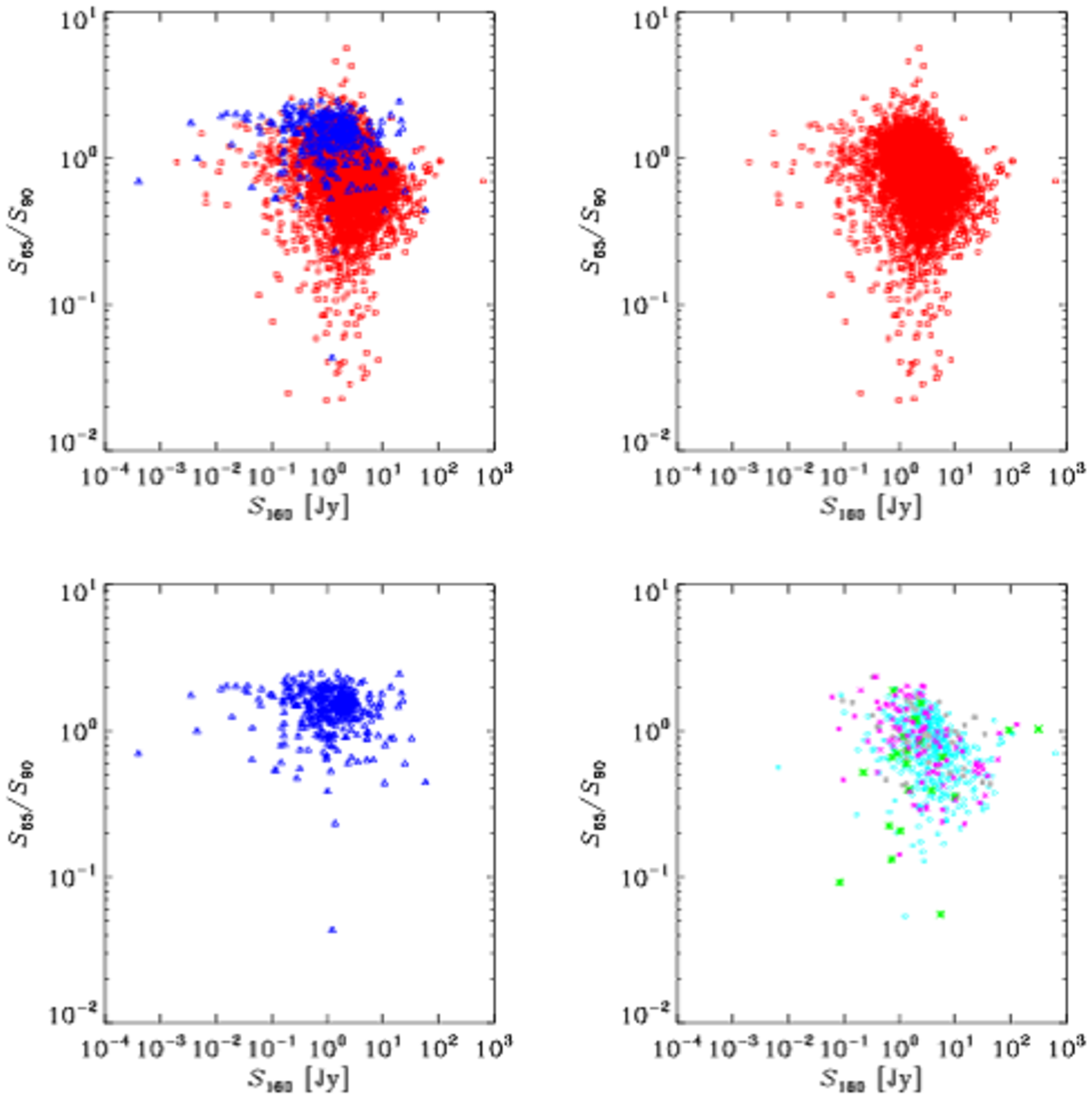}
	\includegraphics[width=0.45\textwidth]{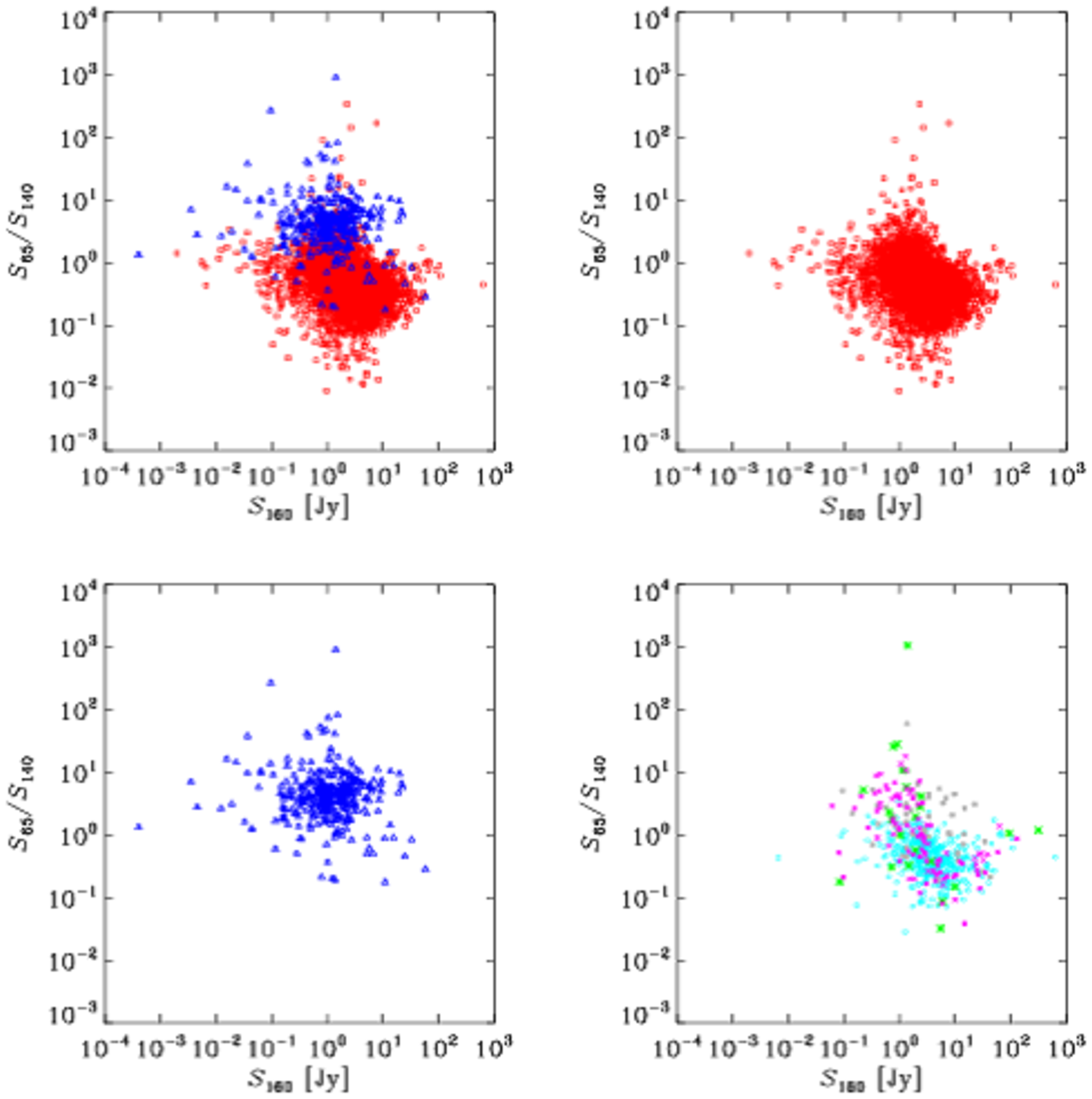}
	\caption{Same as Fig.~\ref{fig:fc1} but for 
	$S_{160}$--$S_{65}/S_{90}$ (four panels on the left side) 
	$S_{160}$--$S_{65}/S_{140}$ (four panels on the right side). 
	} 
	\label{fig:afc9}
\end{figure}
%}

%\onfig{8}{
\begin{figure}[thb]
	\centering
	\includegraphics[width=0.45\textwidth]{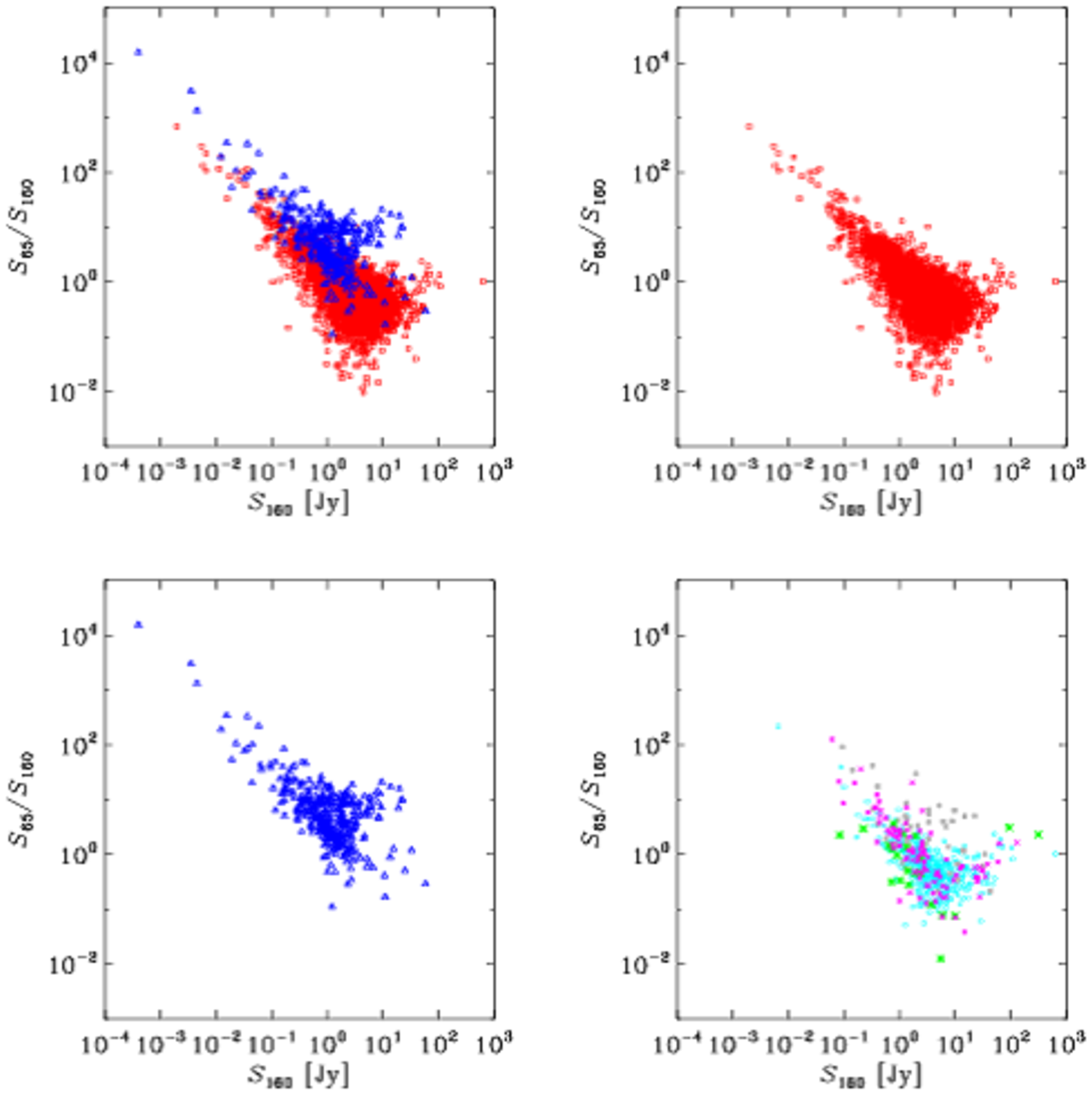}
	\includegraphics[width=0.45\textwidth]{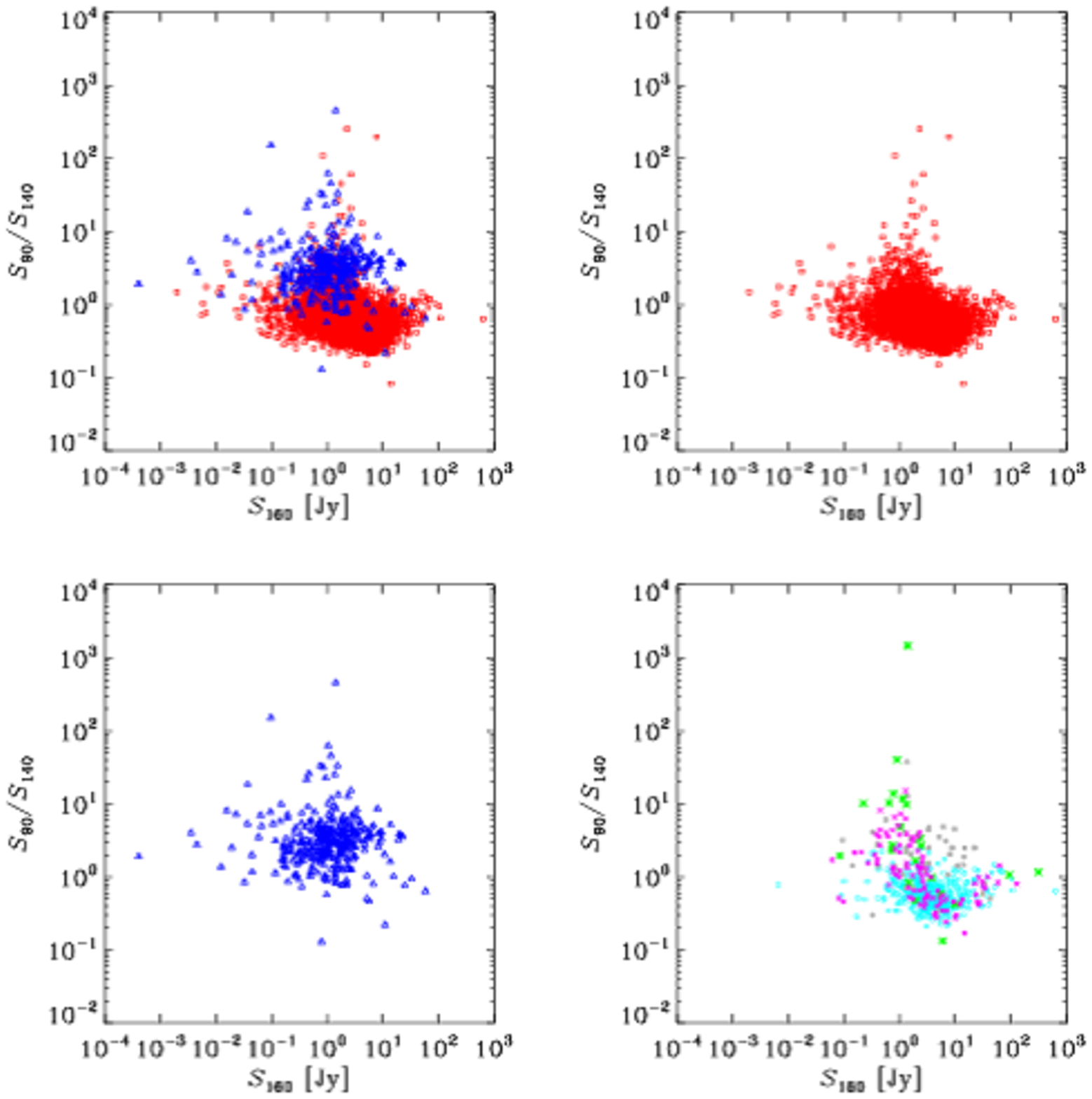}
	\caption{Same as Fig.~\ref{fig:fc1} but for 
	$S_{160}$--$S_{65}/S_{160}$ (four panels on the left side) and
	$S_{160}$--$S_{90}/S_{140}$ (four panels on the right side).  }  
	 \label{fig:afc10}
\end{figure}
%}

%\onfig{9}{
\begin{figure}[thb]
	\centering
	\includegraphics[width=0.45\textwidth]{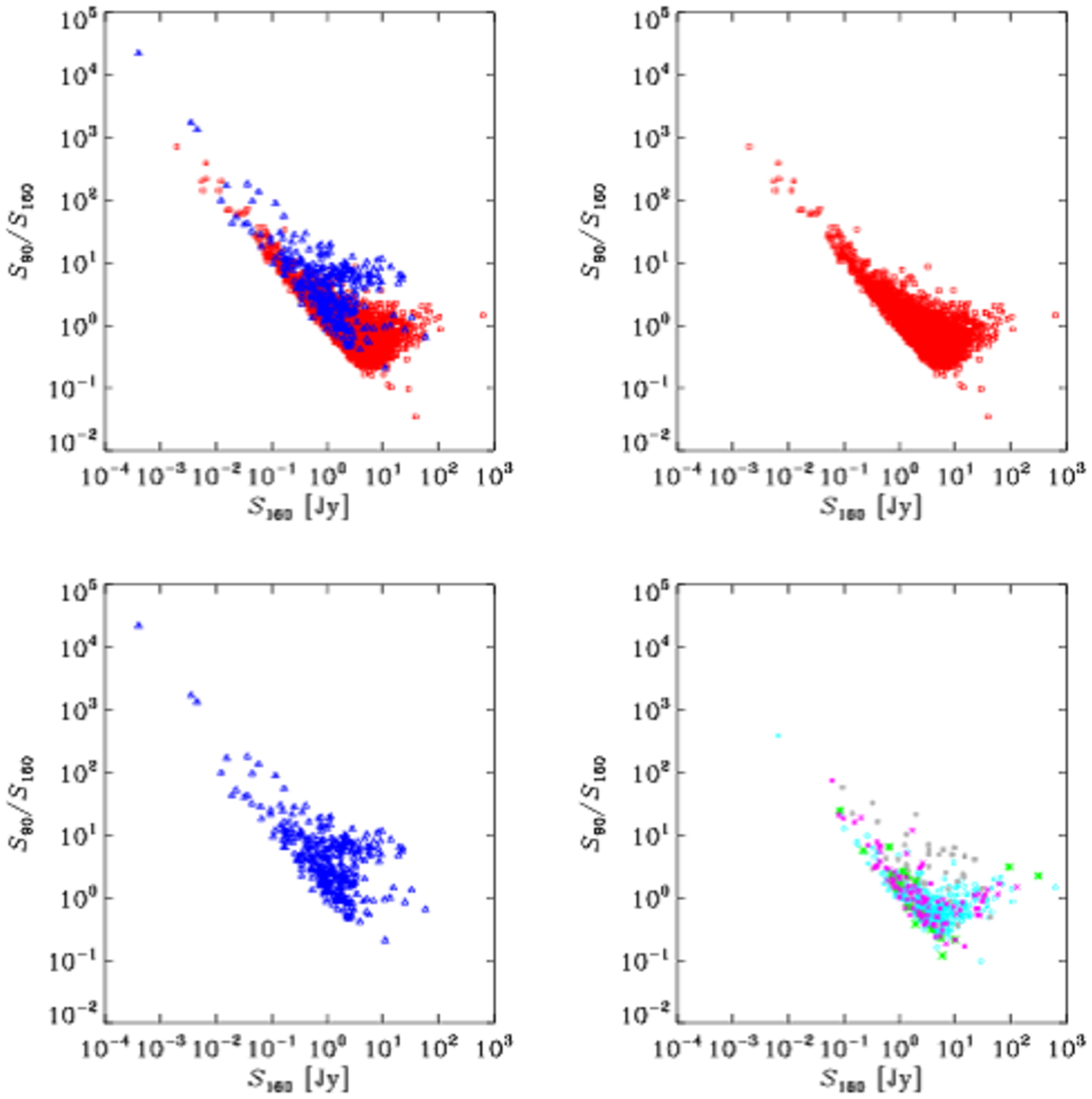}
	\includegraphics[width=0.45\textwidth]{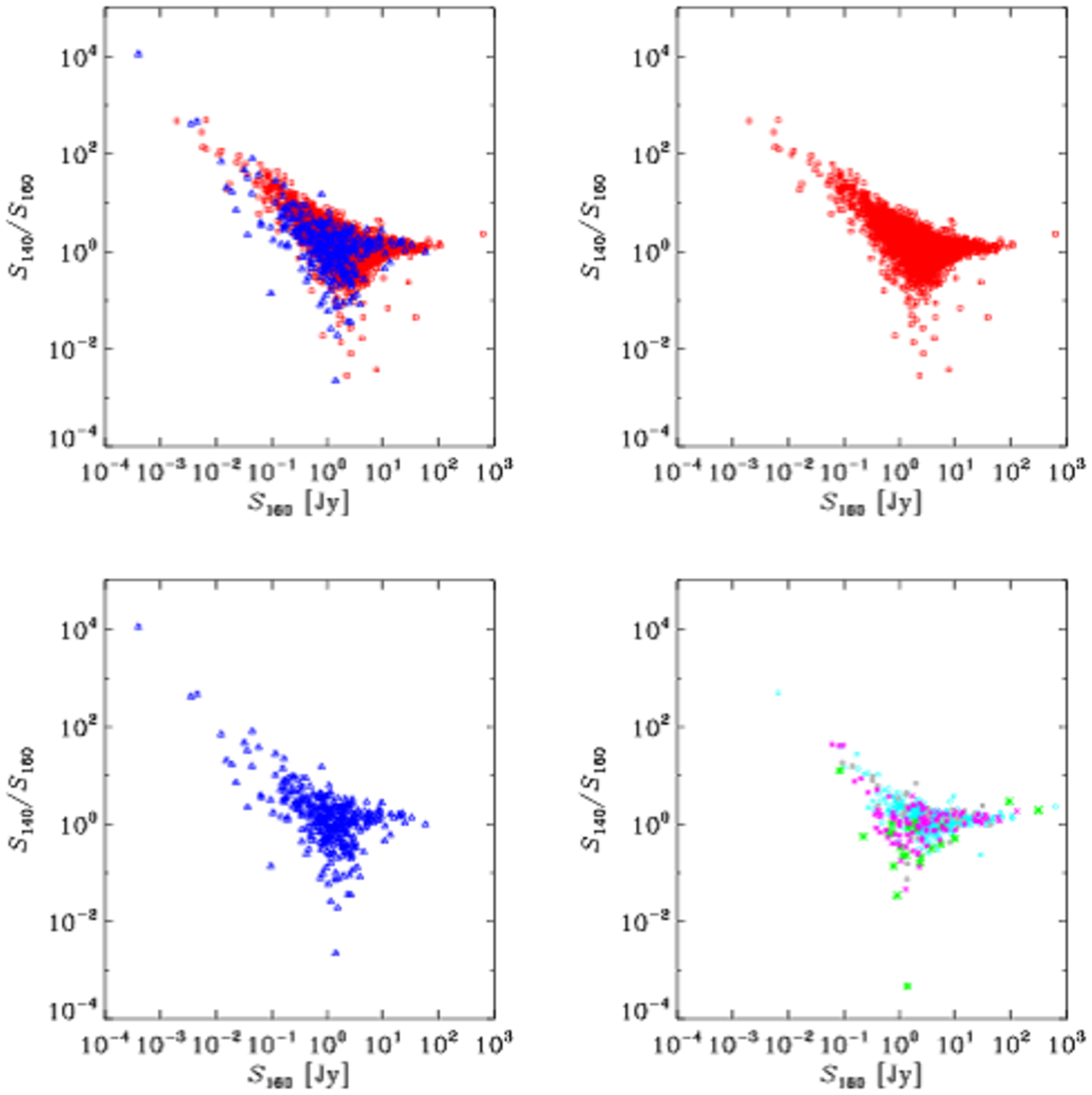}
	\caption{Same as Fig.~\ref{fig:fc1} but for 
	$S_{160}$--$S_{90}/S_{160}$ (four panels on the left side) and
	$S_{160}$--$S_{140}/S_{160}$ (four panels on the right side).  }  
	 \label{fig:afc11}
\end{figure}
%}

\clearpage

\section{Other color-color diagrams}

%\onfig{10}{
\begin{figure}[thb]
	\centering
	\includegraphics[width=0.45\textwidth]{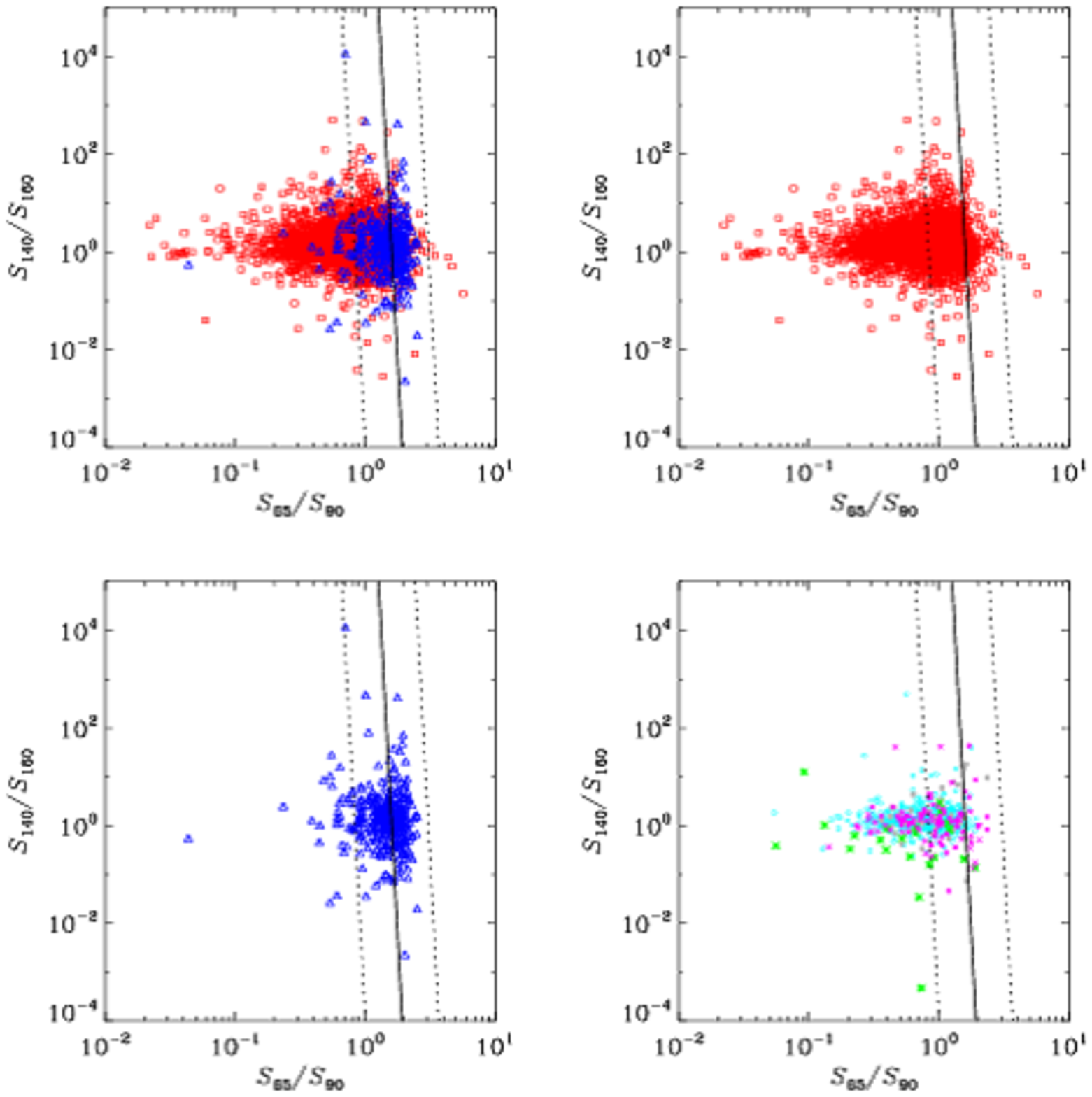}
\includegraphics[width=0.45\textwidth]{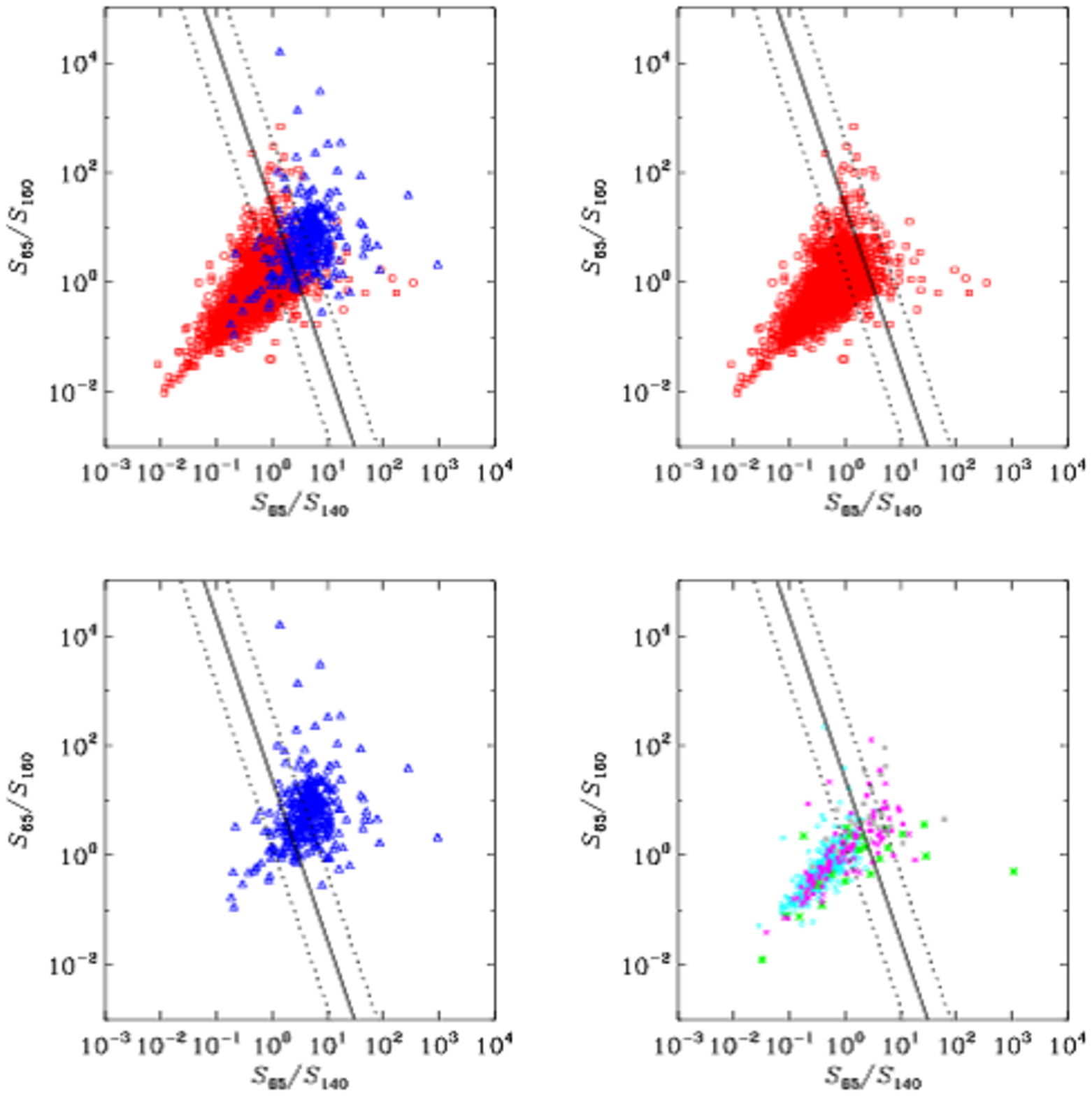}
	\caption{Same as Fig.~\ref{fig:cc1} but for 
$S_{140}/S_{160}$--$S_{65}/S_{90}$ (four panels on the left side) and
$S_{65}/S_{160}$--$S_{65}/S_{140}$ (four panels on the right side).}  
	 \label{fig:cc5}
\end{figure}
%}

%\onfig{11}{
\begin{figure}[thb]
	\centering
	\includegraphics[width=0.45\textwidth]{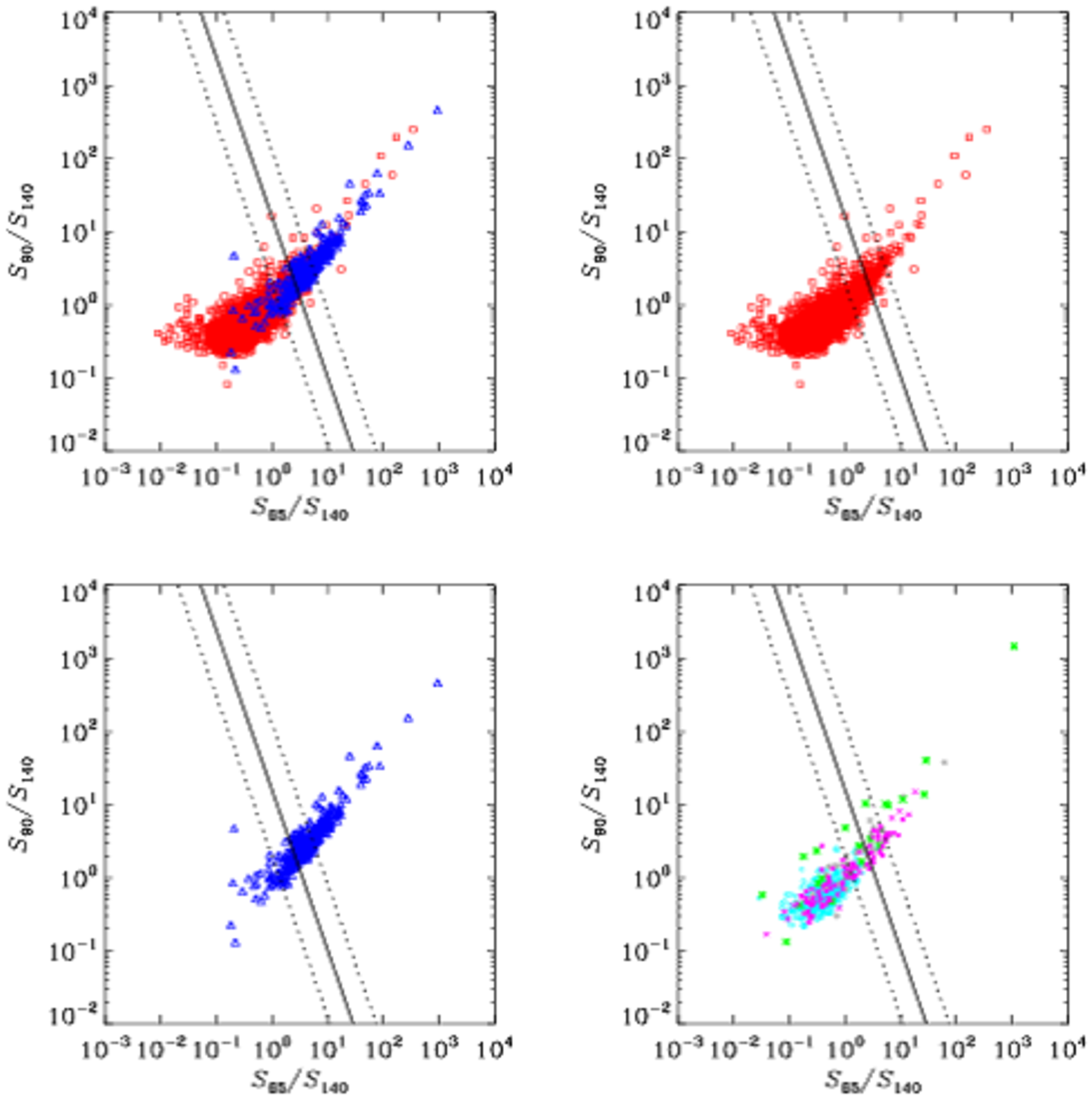}
        \includegraphics[width=0.45\textwidth]{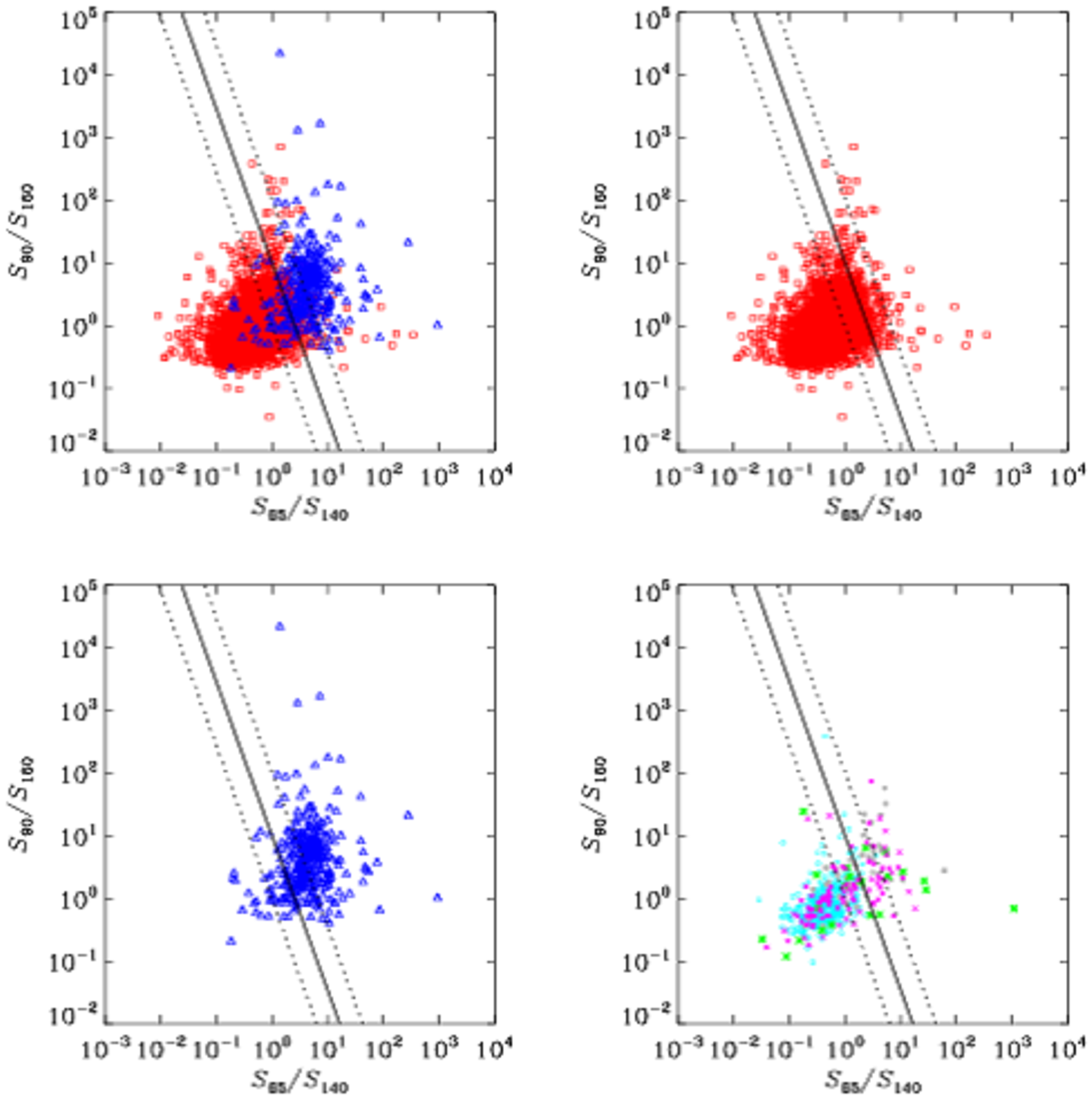}
	\caption{Same as Fig.~\ref{fig:cc1} but for 
$S_{90}/S_{140}$--$S_{65}/S_{140}$ (four panels on the left side) and
$S_{90}/S_{160}$--$S_{65}/S_{140}$ (four panels on the right side).}  
	\label{fig:cc7}
\end{figure}
%}

%\onfig{12}{
\begin{figure}[thb]
	\centering
	\includegraphics[width=0.45\textwidth]{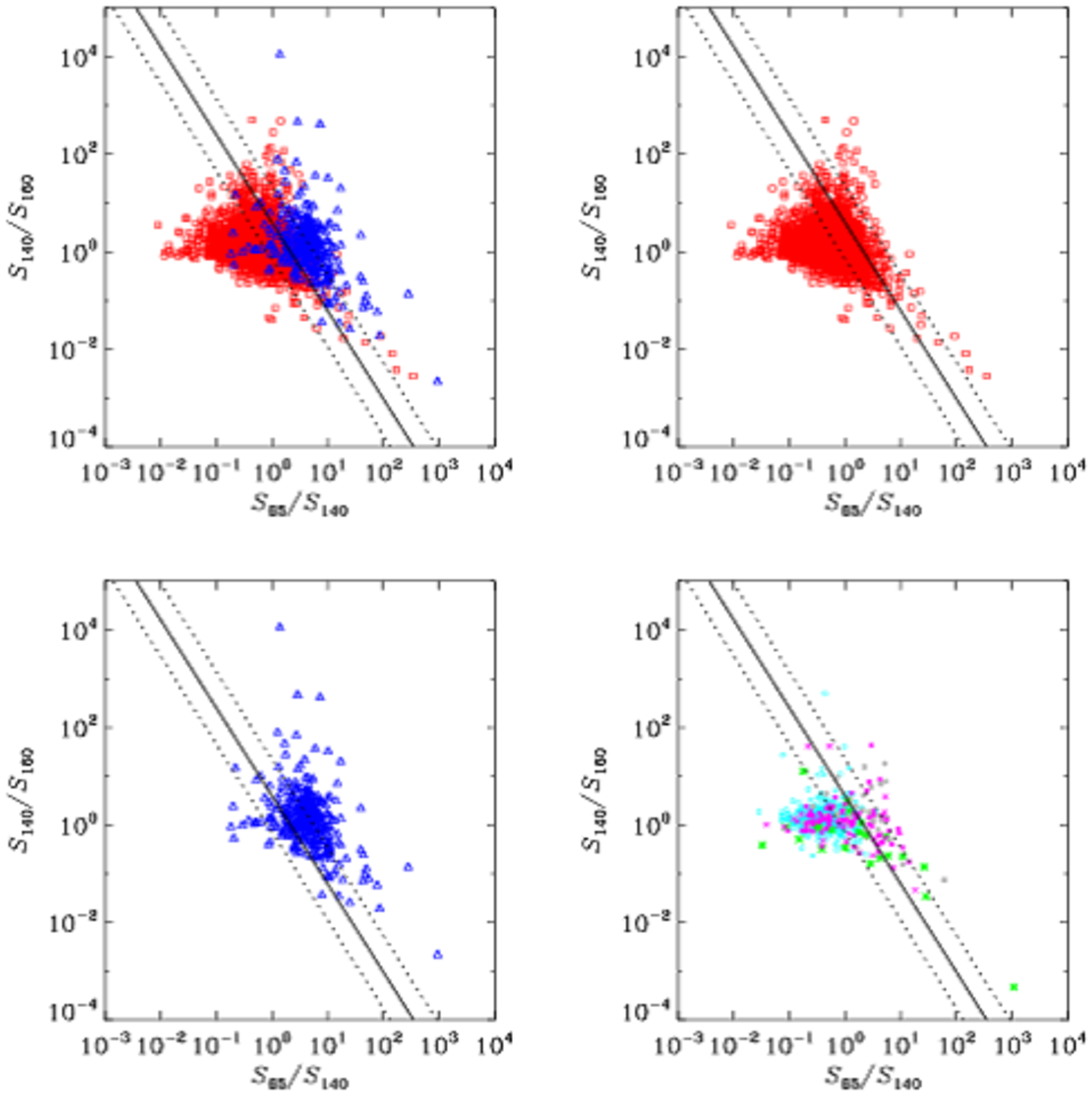}
\includegraphics[width=0.45\textwidth]{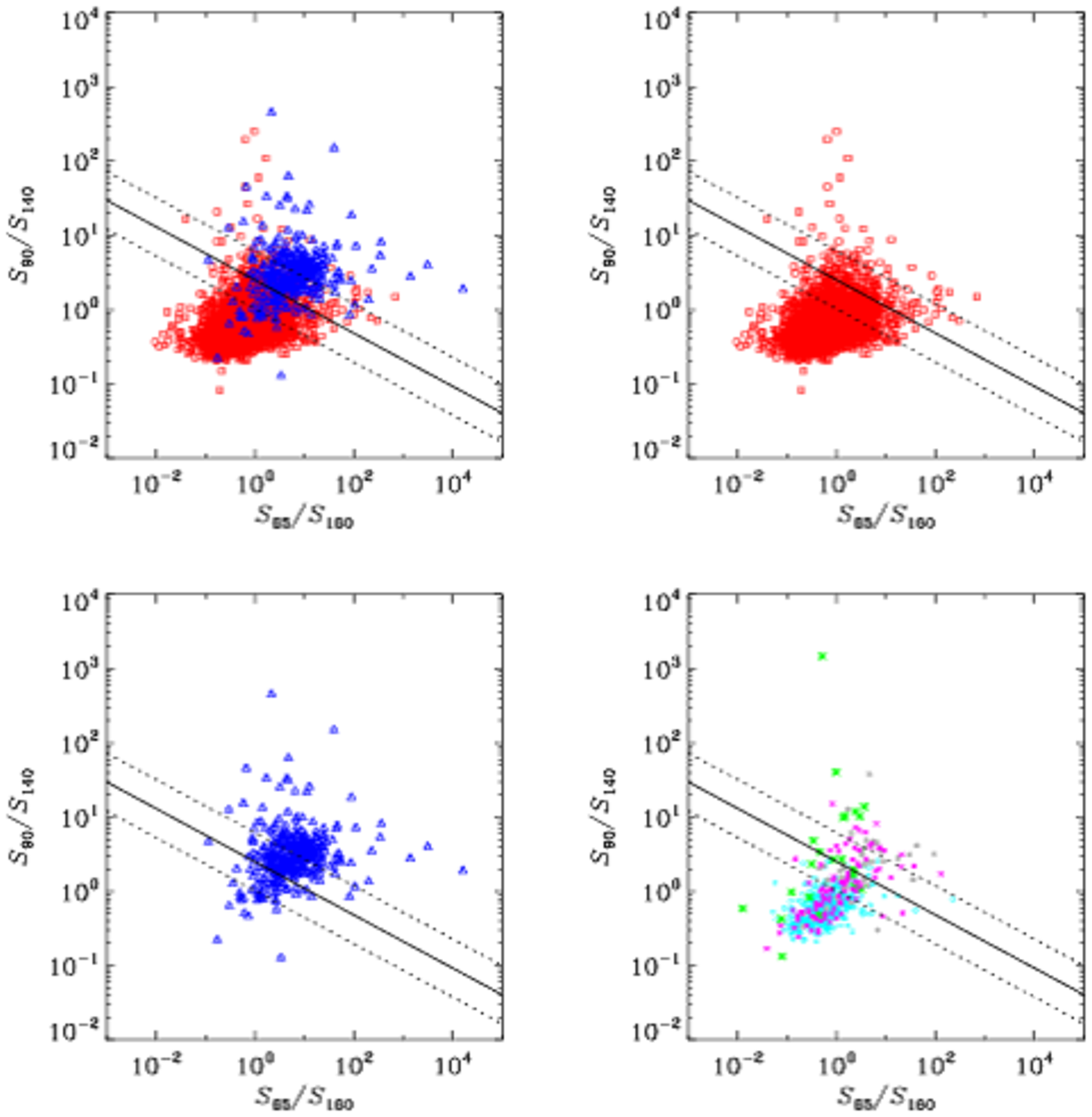}
	\caption{Same as Fig.~\ref{fig:cc1} but for 
$S_{140}/S_{160}$--$S_{65}/S_{140}$ (four panels on the left side) and
$S_{90}/S_{140}$--$S_{65}/S_{160}$ (four panels on the right side).}  
	\label{fig:cc9}
\end{figure}
%}

%\onfig{13}{
\begin{figure}[thb]
	\centering
	\includegraphics[width=0.45\textwidth]{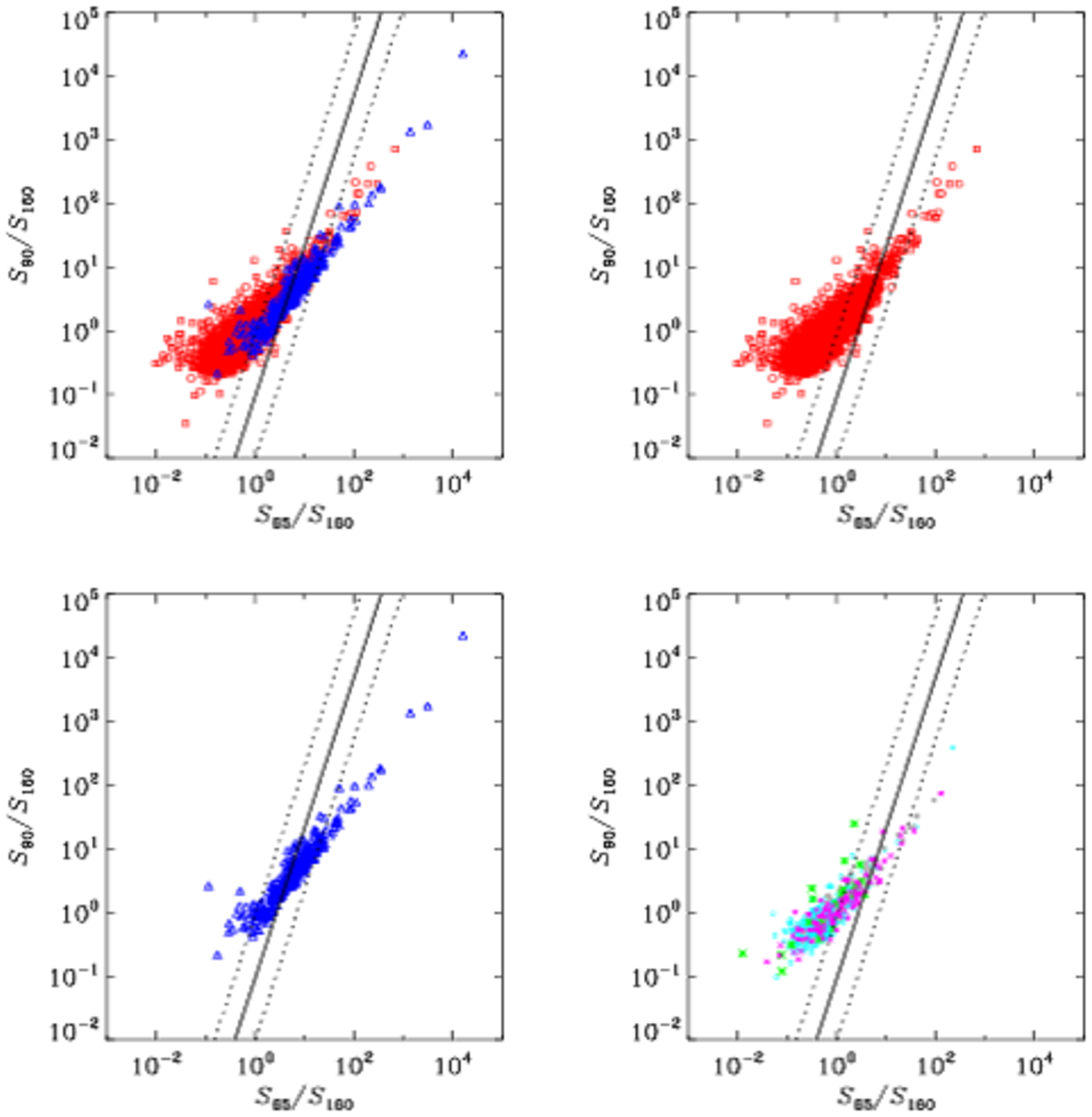}
\includegraphics[width=0.45\textwidth]{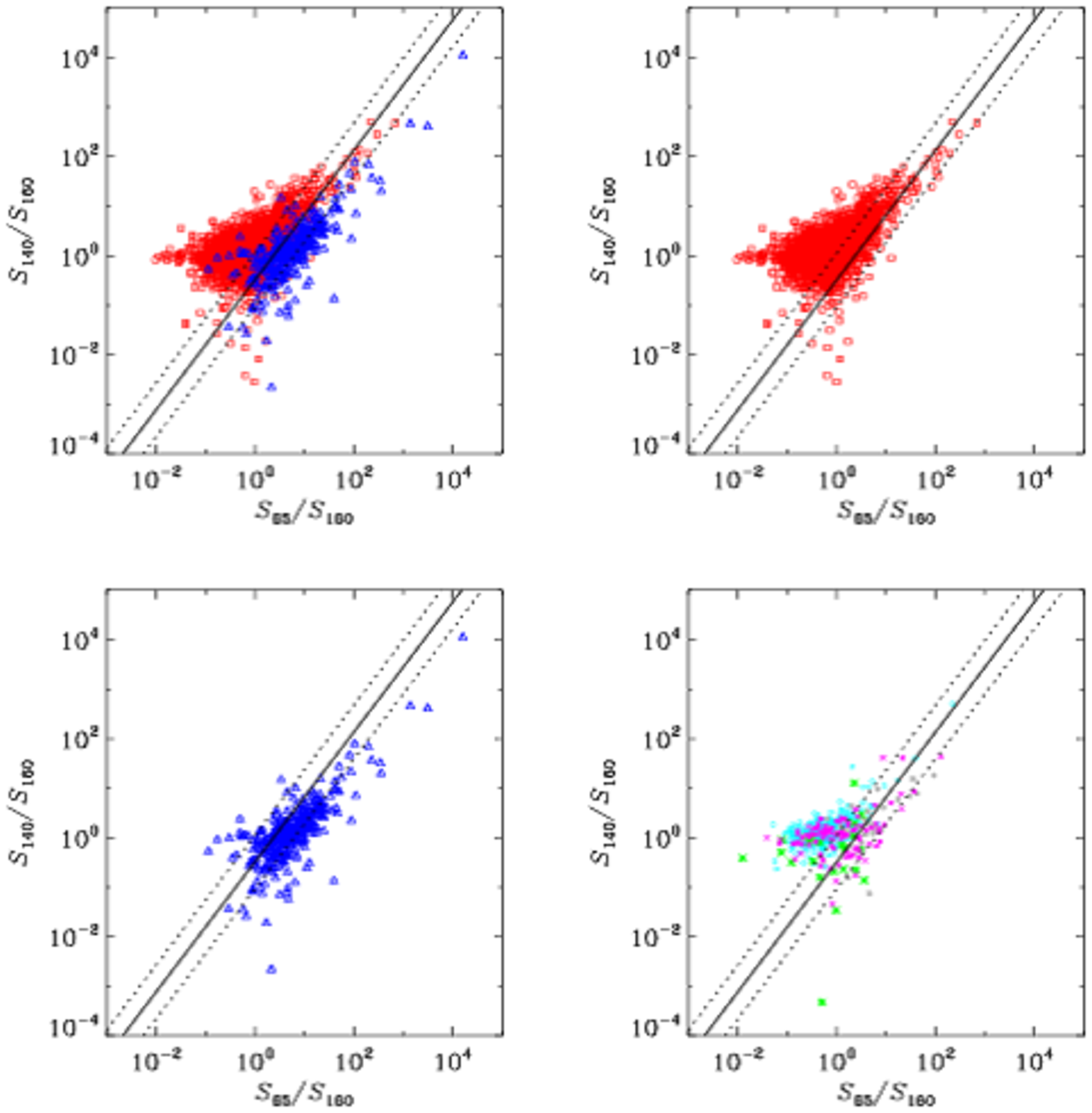}
	\caption{Same as Fig.~\ref{fig:cc1} but for 
$S_{90}/S_{160}$--$S_{65}/S_{160}$ (four panels on the left side) and
$S_{140}/S_{160}$--$S_{65}/S_{160}$ (four panels on the right side).}  
	 \label{fig:cc11}
\end{figure}
%}

%\onfig{13}{ 
\begin{figure}[thb]
	\centering
	\includegraphics[width=0.45\textwidth]{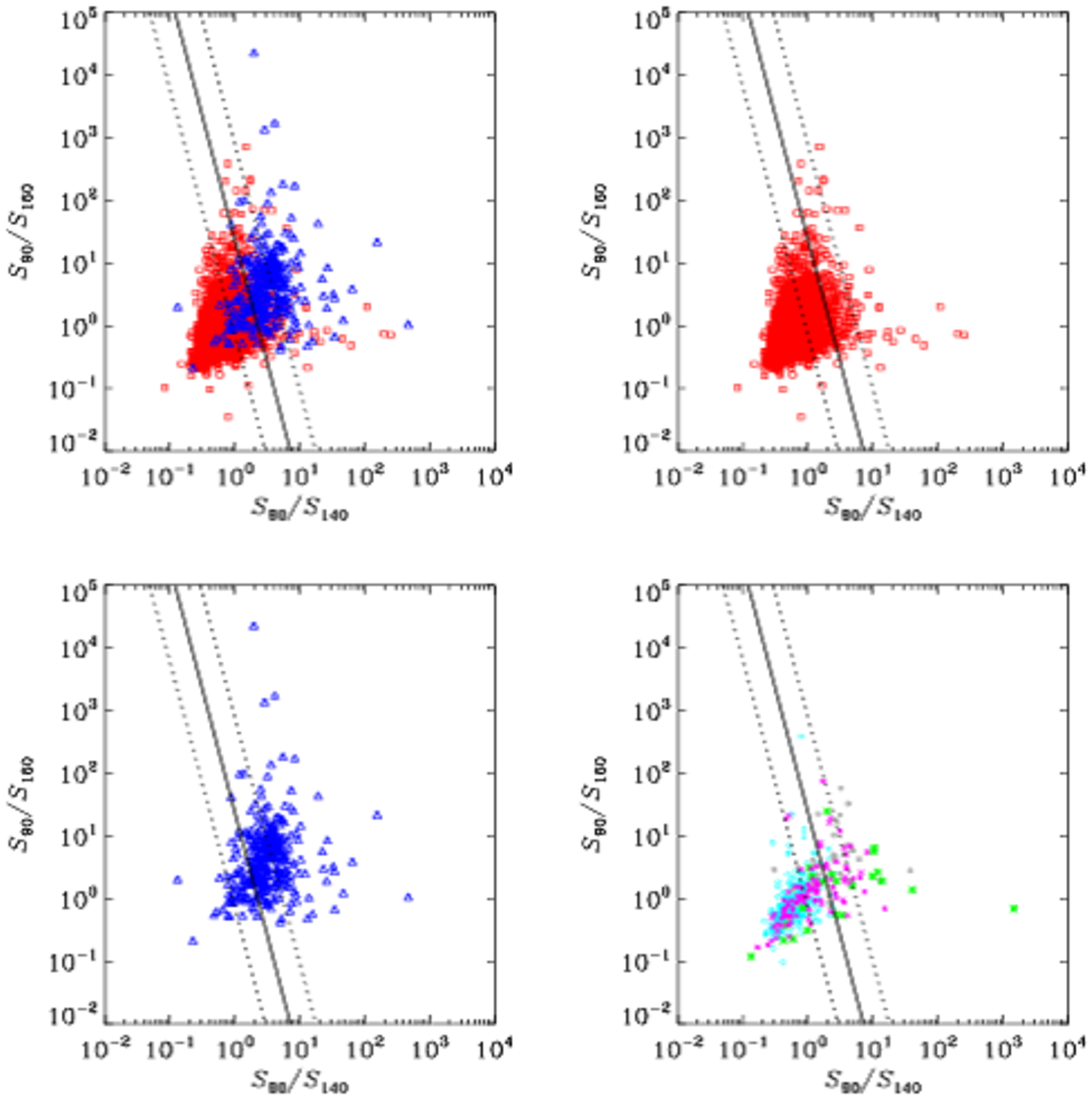}
\includegraphics[width=0.45\textwidth]{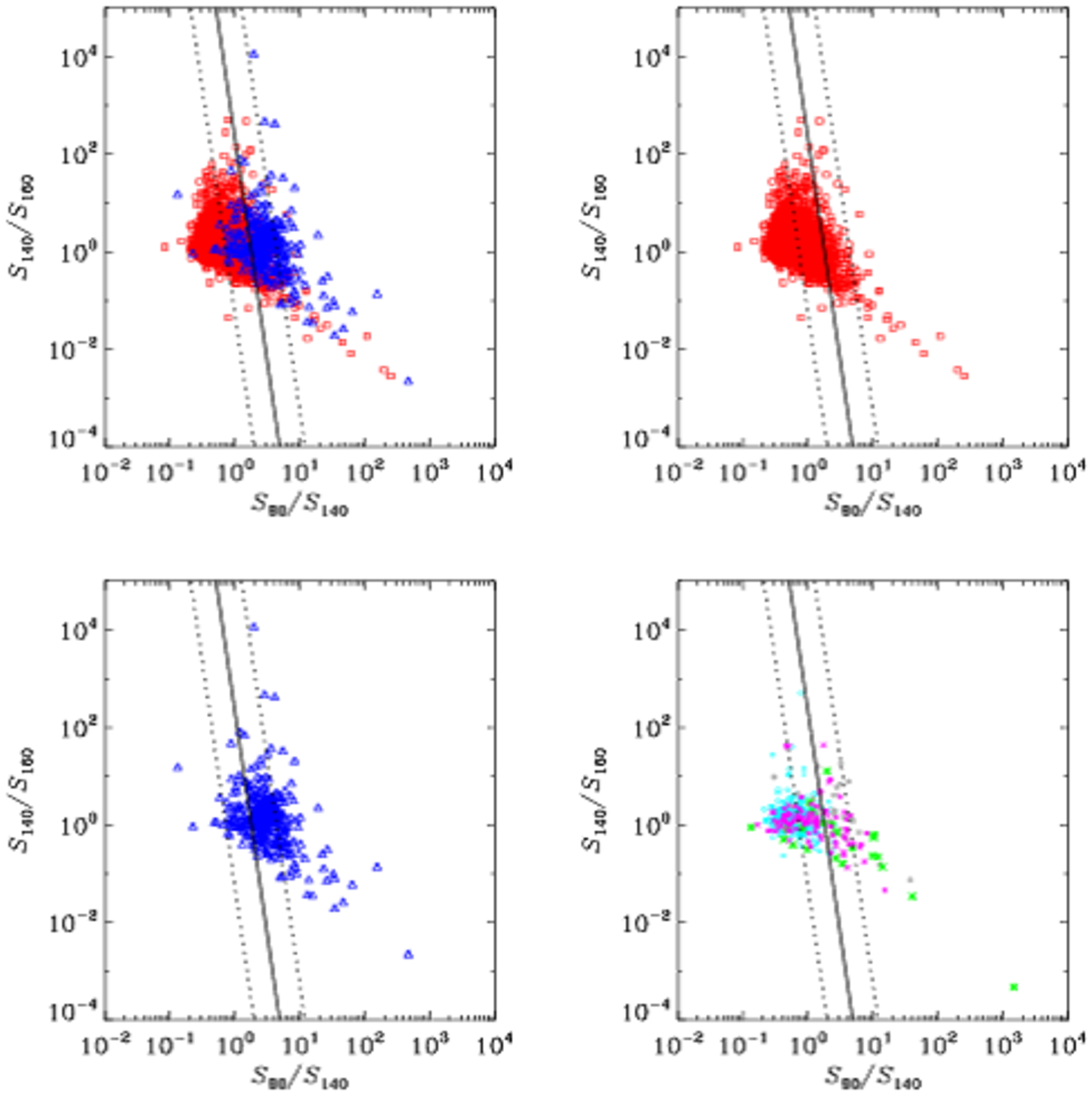}
	\caption{Same as Fig.~\ref{fig:cc1} but for 
$S_{90}/S_{160}$--$S_{90}/S_{140}$ (four panels on the left side) and
$S_{140}/S_{160}$--$S_{90}/S_{140}$ (four panels on the right side).}  
	 \label{fig:cc13}
\end{figure}
%}

%\onfig{14}{
\begin{figure}[thb]
	\centering
	\includegraphics[width=0.45\textwidth]{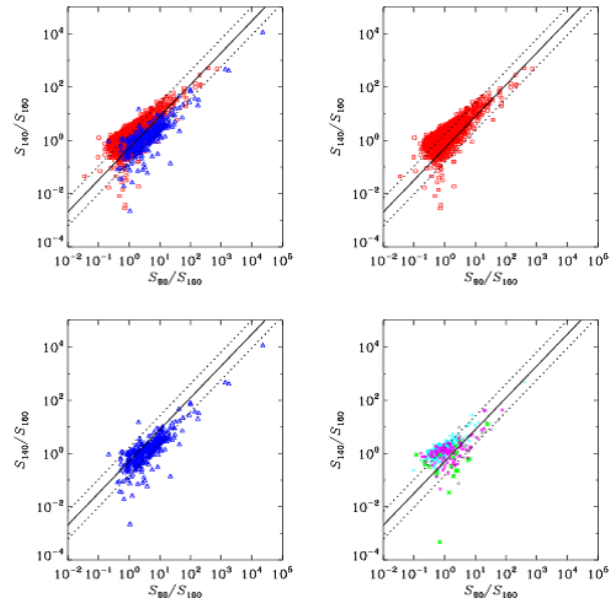}
	\caption{Same as Fig.~\ref{fig:cc1} but for $S_{140}/S_{160}$--$S_{90}/S_{160}$.}  
	 \label{fig:cc15}
\end{figure}
%}


\begin{thebibliography}{}
\bibitem[Buat et al.(2007a)]{buat07a} Buat, V., Takeuchi, T.~T., Iglesias-P\'aramo, J., et al. 2007a, ApJS, 173, 404
\bibitem[Buat et al.(2007b)]{buat07b} Buat, V.,  Marcillac, D., Burgarella, D. et al.  2007b, A\&A, 469, 19
\bibitem[Buat et al.(2008)]{buat08} Buat, V., Boissier, S., Burgarella, D. et al. 2008, A\&A, 483, 107
\bibitem[Caputi et al.(2007)]{caputi07} Caputi, K, Lagache, G., Yan, L. et al. 2007, ApJ, 660, 97
\bibitem[Le Floc'h et al.(2005)]{lefloch05} Le Floc'h, E., Papovich, C., Dole, H.  2005, ApJ, 632, 169 
\bibitem[Ma{\l}ek et al.(2010)]{malek10} Ma{\l}ek, K., Pollo, A., Takeuchi, T.~T., Bieniaz, P., Shirahata, M., Matsuura, S., \& Kawada, M.
	2010, \aap, in press (arXive: astro-ph/0911.5598)
\bibitem[Reddy et al.(2008)]{reddy08} Reddy, N.~A.,  Steidel, C.~C., Pettini, M. et al.  2008, ApJS, 175, 48
\bibitem[Kawada et al.(2007)]{kawada07} Kawada, M., et al.\ 2007, \pasj, 59, 389 
\bibitem[Murakami et al.(2007)]{murakami07} Murakami, H., et al.\ 2007, \pasj, 59, 369
\bibitem[Neugebauer et al.(1984)]{neugebauer84} Neugebauer, G., et al.\ 1984, ApJ, 278, L1 
\bibitem[Onaka et al.(2007)]{onaka07} Onaka, T., et al.\ 2007, PASJ, 59, S401
\bibitem[Schlegel et al.(1998)]{schlegel98} Schlegel, D.~J., Finkbeiner, D.~P., \& Davis, M.\ 1998, \apj, 500, 525 
\bibitem[Takeuchi et al.(1999)]{takeuchi99} 
{
  Takeuchi, T.~T., Hirashita, H., Ohta, K., Hattori, T.~G., Ishii, T.~T., 
  \& Shibai, H.\ 1999, \pasp, 111, 288 
}
\bibitem[Takeuchi et al.(2005)]{takeuchi05a} Takeuchi, T.~T., Buat, V., Burgarella, D. 2005, A\&A, 440, L17
\bibitem[Takeuchi et al.(2010)]{takeuchi10} Takeuchi, T.~T., Buat, V., Heinis, S., Giovannoli, E., Yuan, F.-T., Iglesias-P\'aramo, J.,
	Murata, K.~L., \& Burgarella, D. 2010, \aap, in press (arXiv: astro-ph/Fig.~0912.5051)
\bibitem[Walker et al.(1989)]{walker89} Walker, H.\ J., Cohen, M., Volk, K., Wainscoat, R.\ J., \& Schwartz, D.\ E.\ 1989, AJ, 98, 2163
\bibitem[Yamamura et al.(2009)]{yamamura09} Yamamura, I., et al.\ 2009 in AKARI, a light to illuminate the misty Universe, 
eds.\ T.\ Onaka, G.\ White, T.\ Nakagawa, I.\ Yamamura, ASP Conf.\ Ser., in press
\bibitem[Yamamura et al.(2008)]{yamamura08} Yamamura, I., Fukuda, Y., \& Makiuti, S.\ 2008, 
AKARI/FIS All-Sky  Survey Bright Source Catalogue Version $\beta$-1 Release Note (Rev.\ 1)
\end{thebibliography}
\end{document}